\documentclass{aa}

\usepackage{graphicx}
\usepackage{xcolor}
\usepackage{txfonts}
\usepackage[draft]{hyperref}
\begin{document}
   \title{Age-dating the young open cluster UBC\,1 with g-mode asteroseismology, gyrochronology, and isochrone fitting \thanks{The full Table 1 is only available in electronic form at the CDS via anonymous ftp to cdsarc.u-strasbg.fr (130.79.128.5) or via http://cdsweb.u-strasbg.fr/cgi-bin/qcat?J/A+A/}}


   \author{D. J. Fritzewski\inst{1}
          \and
          T. Van Reeth\inst{1}
          \and
          C. Aerts\inst{1,2,3}
          \and
          J. Van Beeck\inst{1}
          \and
          S. Gossage\inst{4}
          \and
          G. Li\inst{1}
          }

   \institute{Institute of Astronomy, KU Leuven, Celestijnenlaan 200D, 3001, Leuven, Belgium\\
   	\email{dario.fritzewski@kuleuven.be}
      \and
      Department of Astrophysics, IMAPP, Radboud University Nijmegen, PO Box 9010, 6500 GL Nijmegen, The Netherlands
      \and
      Max Planck Institut für Astronomie, Königstuhl 17, 69117 Heidelberg, Germany
      \and
      Center for Interdisciplinary Exploration and Research in Astrophysics (CIERA), Northwestern University, 2145 Sheridan Road, Evanston, IL 60208, USA
   }

   \date{}


  \abstract
   {}
   {UBC\,1 is an open cluster discovered in \emph{Gaia} data and located near the edge of the Transiting Exoplanet Survey Satellite's (TESS) continuous viewing zone. We aim to provide age constraints for this poorly studied open cluster from the combination of gravity-mode (g-mode) asteroseismology, gyrochronology, and isochrone fitting.}
   {We established the members of UBC\,1 from a spatial-kinematic filtering and estimate the cluster age and its parameters. Firstly, we fitted rotating isochrones to the single star cluster sequence. Secondly, using TESS time-series photometry, we explored the variability of the upper main sequence members and identified potential g-mode pulsators. For one star, we found a clear period spacing pattern that we used to deduce the buoyancy travel time, the near-core rotation rate, and an asteroseismic age. For a third independent age estimate, we employed the rotation periods of low-mass members of UBC\,1.}
   {Based on isochrone fitting, we find $\log t = 8.1\pm0.4$, where the large uncertainty occurs because UBC\,1 does not host evolved stars. From asteroseismology of one g-mode pulsator, we find a constrained age of $\log t= 8.24^{+0.43}_{-0.14}$. From gyrochronology based on 17 cool star cluster members, we estimate $\log t = 8.35^{+0.16}_{-0.25}$. Combined, all three methods lead to a consistent age in the range of $150-300$\,Myr.}
   {Our results show that even a single cluster member with identified g modes can improve age-dating of young open clusters. Combining gyrochronology of low-mass members with asteroseismology of intermediate-mass members is a powerful tool for young open cluster modelling, including high-precision age-dating.}

   \keywords{Asteroseismology -- stars: variables: general -- stars: rotation -- open clusters and associations: individual: UBC\,1 -- techniques: photometric}

\titlerunning{Age-dating the open cluster UBC\,1 with asteroseismology, gyrochronology, and isochrones}
\authorrunning{Fritzewski et al.}

   \maketitle
%

\section{Introduction}
Asteroseismology is a relatively recent method to perform stellar modelling, offering precise global parameters, as well as estimates
of the internal physics of stars \citep[see e.g.][for recent reviews on its application to various types of stars and evolutionary stages]{HekkerJCD2017,GarciaBallot2019,Corsico2019,Aerts2021}. It is
based on observed and identified stellar pulsation modes and their frequencies. Yet, most asteroseismic properties of stellar interiors are inferred from stellar models, which have to be calibrated with stars of well-known properties. Among the best known calibrators in stellar astrophysics are stars in open clusters because their common formation history and large mass range provides tight (initial birth) constraints on any model input physics.

Here, we are concerned with young open clusters to assess the internal physics of their members in the core-hydrogen-burning phase of evolution. Such main-sequence stars in young open clusters have been the target of asteroseismic studies from the ground for a long time \citep[e.g.][]{1972ApJ...176..367B, 2000A&A...358..287M, Saesen2013, Mozdzierski2019}. However, the observational restrictions limited the yield and most progress in ground-based asteroseismic modelling of main-sequence stars was achieved for bright field stars \citep[e.g.][to list just a few among many studies covering low- to high-mass dwarfs]{Matthews1999,Pijpers2003,Kervella2004,Bedding2006,Bazot2007,Briquet2007,Garcia2009,Daszynska2010}.

Both asteroseismology and open cluster astrophysics gained new momentum with the space missions \emph{Kepler} \citep{2010Sci...327..977B}, the Transiting Exoplanet Survey Satellite (TESS, \citealt{2014SPIE.9143E..20R}), and \emph{Gaia} \citep{2016A&A...595A...1G}. We refer to the reviews by \cite{Aerts2021} and \cite{2022Univ....8..111C}, for extensive discussions of space asteroseismology of field stars and space astrometry of clusters, respectively. While \emph{Kepler} and TESS provide unprecedented continuous photometric time-series that allow for high-precision frequency determination of oscillating stars, \emph{Gaia} delivers precise astrometric, photometric, and spectroscopic parameters for the majority of stars in our Galactic neighbourhood. By combining the data from \emph{Gaia} and \emph{Kepler} or TESS, we are now in the position to test stellar models, including those of main-sequence pulsators, with open cluster members more firmly than ever before and thus providing passageways towards improving these models.

For this work, we are mainly concerned with gravity-mode (g-mode) pulsators in the main-sequence stage of their evolution. Specifically, we work with $\gamma$\,Doradus ($\gamma\,$Dor) pulsators, which are intermediate-mass dwarfs.
The \emph{Kepler} and \emph{Gaia} missions led to the discovery \citep{2013A&A...556A..52T, 2020MNRAS.491.3586L, 2023A&A...674A..36G}, description \citep{2015ApJS..218...27V,2016A&A...593A.120V,Ouazzani2017,2018A&A...618A..24V,Mombarg2020,Ouazzani2020,Saio2021,2023A&A...672A.183A}, and detailed asteroseismic modelling \citep{Kurtz2014,Saio2015,SchmidAerts2016,2019MNRAS.485.3248M,Mombarg2021} of these stars, while enabling the development of a broader theoretical underpinning on angular momentum and chemical element transport \citep{Augustson2019,Ouazzani2019,Aerts2019,Augustson2020,
Park2020,Park2021,Prat2021,Dandoy2023,
Mombarg2023}. However, most of this work was carried out on field stars.

Given the ages and distances of the few open clusters in the \emph{Kepler} field, open cluster asteroseismology has mainly focussed on red giants with solar-like oscillations \citep{Basu2011,2011A&A...530A.100H}. (Pre-)main sequence pulsators in open clusters were observed with K2 \citep{2015MNRAS.454.2606R, 2016MNRAS.463.2600L, 2020AJ....159...96S} and lately with TESS \citep[e.g.][]{2021MNRAS.502.1633M, 2023ApJ...946L..10B, 2023A&A...674A.146P}. These studies showed that asteroseismology can constrain the ages of open clusters. However, the studies mostly focussed on pressure mode (p-mode) pulsations in $\delta$\,Sct-type stars located in the classical instability strip. In this work, we exploit the potential of g-mode asteroseismology to age-date a barely studied open cluster and cross-calibrate the asteroseismic age with the ages derived from other methods.

One reason why most asteroseismic studies on open cluster stars with TESS focus on $\delta$\,Sct-type pulsators is its observing mode. Although it provides time-series photometry with a nearly all-sky coverage, the 27\,d-coverage by its sectors hinders deep asteroseismic exploration of the data when the beating patterns of multi-periodic oscillations are longer than this time base. Moreover, due to the short sector baseline, individual frequencies in the g-mode regime ($0.5\lesssim f_\mathrm{puls} \lesssim 5$\,d$^{-1}$) are not resolved because the frequency resolution is inversely proportional to that baseline. Fortunately, the TESS observations include a continuous viewing zone (CVZ) in each hemisphere which is monitored for 352\,d, hence providing long-baseline time-series photometry that enables precise frequency determination for multi-periodic low-frequency pulsations.

One of the few open clusters with a large number of observed TESS sectors (i.e. in or on the edge of the CVZ) is the recently discovered UBC\,1 (others include the tidal tails of NGC\,2516 which are treated in a separate parallel study, Li et al., in prep.). \cite{2018A&A...618A..59C} discovered this open cluster in \emph{Gaia}~DR2 data with an unsupervised clustering algorithm. As \cite{2018A&A...618A..59C} note the position of this open cluster matches the previously described RSG\,4 \citep{2016A&A...595A..22R}, yet the mean proper motion and the distance of the clusters do not agree. The Milky Way Star Cluster catalogue \citep{2014A&A...568A..51S} also includes an entry for an open cluster near the position of UBC\,1. However, MWSC\,5373 is located at a distance of 1.6\,kpc (compared to 320\,pc for UBC\,1) and can therefore not be considered the same cluster.

UBC\,1 has not been analysed in a dedicated study, yet it is included in other large-scale, \emph{Gaia}-based open cluster studies. \cite{2019AJ....158..122K} include UBC\,1 in their list of stellar clusters and streams as Theia\,520 with an age of 182\,Myr and \cite{2019JKAS...52..145S} list it as UPK\,134 with similar astrometric parameters albeit with an estimated age of 525\,Myr\footnote{We note that in \cite{2019JKAS...52..145S}, UBC\,1 is confused with MWSC\,5373 based on the sky position despite their different distances.}. \cite{2020A&A...640A...1C} revisit the collection of UBC clusters, update the cluster parameters slightly and estimate the age to be 70\,Myr with a reddening of $A_V=0.35$\,mag. We note that the estimated age of RSG\,4 is 350\,Myr \citep{2016A&A...595A..22R}.

We aim to explore the (g-mode) pulsators in UBC\,1 to obtain an asteroseismic age estimate, along with independent age-dating from isochrone fitting based on rotating stellar models and from gyrochronology. After establishing the membership in Sect.~\ref{sec:members}, we find the isochronal age based on \emph{Gaia} photometry in Sect.~\ref{sec:isoage}. Further, we describe the TESS observations and our data reduction, and identify promising cluster pulsators (Sect.~\ref{sec:obs}). Using this information, we estimate the asteroseismic age of UBC\,1 in Sect.~\ref{sec:asteroage}. To refine our age estimate and to exploit the full potential of TESS, we also estimate the gyrochronal age based on cool star rotation periods in Sect.~\ref{sec:gyroage}. Finally, we briefly discuss the different age estimates and come to conclusions in Sect.~\ref{sec:discuss}.

\section{Open cluster membership}
\label{sec:members}
\subsection{Membership}
The membership list of UBC\,1 presented in \cite{2018A&A...618A..59C} and subsequently in \cite{2020A&A...640A...1C} includes 47 members, while \cite{2019JKAS...52..145S} list 94 members. In contrast, \cite{2019AJ....158..122K} found 397 members based on unsupervised clustering. However, many of the listed members may not be true members of the open cluster because the velocity dispersion of the groups found by \cite{2019AJ....158..122K} is often larger than typically found in an open cluster \citep{2021A&A...645A..84M}. In the following, we reanalyse the stars in the field of UBC\,1 to establish a comprehensive, inclusive, yet clean membership.

For our analysis, we followed the kinematic and spatial filtering approach outlined in \cite{2019A&A...621L...3M} and \cite{2021A&A...645A..84M}. This method is rather conservative and traditional, yet thanks to the precision of the \emph{Gaia} data also very accurate. In our implementation, the filtering is a three step process based on (1) proper motions, (2) 3D motion, and (3) spatial distribution. Hence, as in \cite{2021A&A...645A..84M}, we did not use a photometric criterion. We applied these steps to a generous selection of \emph{Gaia}~DR3 \citep{2023A&A...674A...1G} sources centred on the known position of UBC\,1 (see Appendix~\ref{app:adql} for the \emph{Gaia} archive ADQL query).

A prerequisite of calculating the velocity dispersion is knowledge of the bulk motion of the open cluster which we took from  \cite{2018A&A...618A..59C}.
Following \cite{2021A&A...645A..84M}, we first calculated the tangential velocities of each source in km\,s$^{-1}$ and subsequently the velocity dispersion in the sky plane, $\Delta v_\mathrm{2D}$. For the members presented in \cite{2018A&A...618A..59C}, we found a velocity dispersion of 1.6\,km\,s$^{-1}$.
To include all these sources as members, we selected a threshold of 1.7\,km\,s$^{-1}$ and considered all stars with $\Delta v_\mathrm{2D} < 1.7\,\mathrm{km\,s}^{-1}$ as proper motion members.

For the brighter stars in the field, radial velocity measurements are available from \emph{Gaia}~DR3 \citep{2023A&A...674A...5K}. Yet their uncertainties are up to 9\,km\,s$^{-1}$. We only used these data to exclude stars with too divergent radial velocities, defined as $\Delta v_\mathrm{3D} > 10$\,km\,s$^{-1}$. In this way, we found three stars in the membership list of \cite{2018A&A...618A..59C} that exceed this value and were therefore removed.

Although the kinematic filtering was very efficient, we still found many co-moving sources over a large volume. In order to reduce the number of random matches, we applied a spatial density filtering similar to \cite{2019A&A...621L...3M}. For each star, we calculated the distance in pc to the three nearest kinematic members and selected only stars with all neighbours within 10\,pc. This value effectively suppressed all random aggregates of a few stars that could be found in the data but also allowed us to retain members on the outskirts of the open cluster.

\begin{figure}
    \includegraphics[width=\columnwidth]{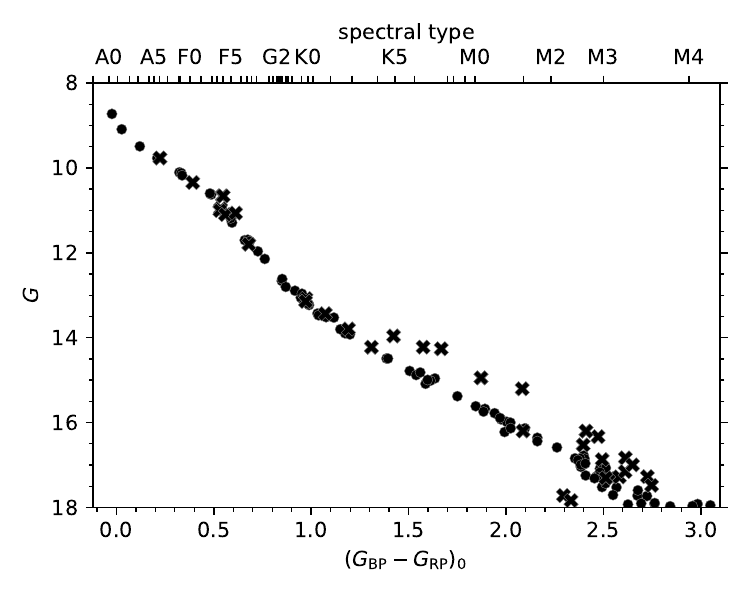}
    \caption{Colour-magnitude diagram of the kinematically and spatially selected members of UBC\,1. Potential binaries are marked with crosses.
}
    \label{fig:CMD}
\end{figure}

In total, we find 132 members of UBC\,1 ranging from spectral type A0 to mid-M (Table~\ref{tab:mem}). Fig.~\ref{fig:CMD} shows the colour-magnitude diagram\footnote{The spectral type axis in this and all subsequent figures is based on \cite{2013ApJS..208....9P} (\url{http://www.pas.rochester.edu/~emamajek/EEM\_dwarf\_UBVIJHK\_colors\_Teff.txt)}} (CMD) of the cluster members. The single star cluster sequence is very clean.
Among the lower-mass stars, we observe a spread in the sequence due to less accurate \emph{Gaia} measurements. The photometric binary sequence is sparsely populated and hosts only cool stars. We note that we did not apply any photometric criteria other than the maximum magnitude of $G=18$, showing again the precision with which co-moving structures can be extracted from the \emph{Gaia} data.

\begin{table*}
    \caption{Membership list for UBC\,1. The full table is available online.}
    \label{tab:mem}
    \begin{tabular}{ccccccc}
        \hline
        \hline
        ID & \emph{Gaia} Designation & RA & Dec. & $\Delta v_\mathrm{2D}$ & $\Delta v_\mathrm{3D}$ & $\varpi$\\
        & & ($\deg$) & ($\deg$) & (km\,s$^{-1}$) & (km\,s$^{-1}$) & (mas)\\
        \hline
    1 & Gaia DR3 2153393527297256448 & 284.78292 & 56.73813 & 1.313 & \dots & 3.276\\
    2 & Gaia DR3 2152926647172254464 & 285.06069 & 56.06692 & 0.864 & 1.456 & 3.241\\
    3 & Gaia DR3 2152897922430896128 & 285.32982 & 55.57524 & 0.698 & \dots & 3.204\\
    4 & Gaia DR3 2140886754330210816 & 285.57663 & 55.49195 & 1.083 & 1.609 & 3.059\\
    5 & Gaia DR3 2153330266721959296 & 285.88908 & 56.62504 & 1.361 & 6.080 & 3.407\\
    6 & Gaia DR3 2153473826005633408 & 285.94592 & 57.38454 & 0.039 & \dots & 3.267\\
    7 & Gaia DR3 2140829579725254144 & 286.01609 & 54.82652 & 1.426 & \dots & 3.153\\
    8 & Gaia DR3 2141307729844679424 & 286.02169 & 56.19164 & 0.948 & 4.177 & 3.049\\
    9 & Gaia DR3 2140868406229943296 & 286.02237 & 55.28755 & 1.463 & \dots & 3.133\\
    \vdots&\vdots&\vdots&\vdots&\vdots&\vdots&\vdots\\
        \hline
    \end{tabular}
\end{table*}

For completeness, we reiterated the membership determination, but this time with the proper motion for RSG\,4 as reported in \cite{2016A&A...595A..22R}. We did not find evidence of a cluster at the given position in velocity space. Hence, we conclude that the detection of RSG\,4 was illusive and we call the open cluster UBC\,1 in this work.

From our analysis, we find a distance of 320\,pc to the cluster core with a mean proper motion of $\mu^*_\alpha=-2.53184$\,mas\,yr$^{-1}$, $\mu_\delta=3.73449$\,mas\,yr$^{-1}$ and a mean radial velocity $v_\mathrm{r}=-23.2$\,km\,s$^{-1}$. Based on the reddening provided by {\em Gaia} \citep{2023A&A...674A..27A}, we expect the cluster to be mostly extinction free. Yet, the numerical value (the average of all members with $T_\mathrm{eff}>4600$\,K) is not well constrained with $E(G_\mathrm{BP}-G_\mathrm{RP})=0.035\pm0.034$\,mag (corresponding to $E(B-V)=0.026$\,mag, \citealt{2018MNRAS.479L.102C}).

The metallicity estimate from \emph{Gaia} \citep{2023A&A...674A..27A} is similarly uncertain with [Fe/H]$=-0.2\pm0.3$. On the other hand, the metallicity determined from APOGEE spectra \citep{2020AJ....160..120J} of five members is consistent with the solar value ([Fe/H]$=0.02\pm0.03$). The stars in common between APOGEE and \emph{Gaia} are offset by $0.15$\,dex \citep[c.f. Fig. 11 in][]{2023A&A...674A..27A}. From the available data and the knowledge that most open clusters in the solar vicinity are also of solar metallicity, we conclude that UBC\,1 is not an outlier.

\subsection{Binarity}
Binarity (multiplicity in general) can influence stellar evolution in many aspects. It can change the star's structure, its chemical mixing, and angular momentum evolution \citep[e.g.][]{2012Sci...337..444S}. All these properties influence stellar pulsations. Hence, knowing potential binaries in our sample is important in the subsequent analyses.

Multiple binarity indicators are available for stars in open clusters. Primarily, we used the information provided in \emph{Gaia} DR3, including the reduced unit weight error (RUWE), spectroscopic binarity indicators, and photometric data. The latter was supplemented by infra-red photometry from WISE \citep{2010AJ....140.1868W, 2011ApJ...731...53M}.

We flagged every star with $\mathrm{RUWE}>1.2$ as a potential binary \citep{2020MNRAS.496.1922B}. An increased RUWE value reflects larger uncertainties in the \emph{Gaia} astrometric solution and can indicate an unresolved binary. \emph{Gaia} DR3 also provides spectroscopic orbits \citep{Arenou2023}
for two upper main sequence members of UBC\,1, which we included in our list of cluster binaries. Unresolved binaries can also be identified from photometry, because they are elevated above the single star cluster main sequence in the colour-magnitude diagram (CMD). As seen from Fig.~\ref{fig:CMD}, photometric binaries can mostly be found among UBC\,1's low-mass members. However in a CMD based on a redder colour ($G-W1$, with W1 from WISE), we find additional outliers redwards of the main sequence (Appendix~\ref{app:binarity}). These unequal-mass binaries have a redder component that contributes more to the combined flux in the redder passband than in the optical \emph{Gaia} passband. The redder component could also be a debris disc as observed around other intermediate mass stars \citep{2006ApJ...653..675S}. Nevertheless, we treated these stars as candidate binaries as a precaution. We note that photometric binaries identified in the optical are also found in the infrared. We therefore flagged all stars found at redder colours than the cluster main sequence in any of the CMDs as photometric binaries.

\emph{Gaia} DR3 also provides a radial velocity amplitude for some sources. Radial velocity variability can be induced by a companion. However, stellar pulsations also cause spectral line-profile variations for dwarfs, leading to radial-velocity changes of tens of km\,s$^{-1}$ depending on the kind of star and the nature of the modes \citep[e.g.][]{1976ApJ...210..163B,Aerts1992,DeCatAerts2002,Aerts2003,Mathias2004, 2011A&A...529L...8K,Aerts2014}. As our targets potentially include such stars, the radial velocity variability cannot be used as a reliable binarity indicator. With future \emph{Gaia} data releases, the epoch radial velocity data will become available and enable the distinction between sources of radial velocity variability based on the observed periodicity and all the combined time-series data.

In total, we find 35 out of 132 members of UBC\,1 (27\,\%) to show signs of potential multiplicity. This percentage is in agreement with studies of other open clusters based on \emph{Gaia} data \citep{2020ApJ...903...93N, 2022ApJ...930...44L, 2023A&A...675A..89D}. We mark all binaries with distinct symbols in the figures.

\section{Isochronal age of UBC\,1}
\label{sec:isoage}
Our membership list includes only main sequence stars, which makes it challenging to estimate the age of this open cluster by means of isochrone fitting. Nevertheless, we employed this method to obtain a first handle on the cluster age and to assess the parameter space for the asteroseismic modelling discussed in Sect.\,5.
Since the asteroseismic models in that Section are constructed with the Modules for Experiments in Stellar Astrophysics software (MESA, \citealt{2011ApJS..192....3P, 2013ApJS..208....4P, 2015ApJS..220...15P}), we relied on the MESA Isochrones and Stellar Tracks (MIST) isochrones as a natural choice \citep{2016ApJS..222....8D, 2016ApJ...823..102C}.

The standard MIST isochrones are provided as non-rotating models and as models rotating at 40\,\% of the critical rate, $v/v_\mathrm{crit}=0.4$ (see \citealt{2016ApJ...823..102C} for the definition of $v_\mathrm{crit}$). As intermediate-mass  stars are typically fast rotators and rotation is known to affect
main-sequence turn-off morphologies of young clusters \citep[e.g.][]{2018ApJ...869..139C, 2018ApJ...863L..33M,2018MNRAS.480.1689K}, we employed the
generalised MIST isochrones computed by \cite{2019ApJ...887..199G} (see also \citealt{gossage_seth_2019_8008601}) to perform the isochrone fitting.
These cover rotation rates ranging from $0\leq v/v_\mathrm{crit}\leq0.9$ in steps of 0.1. Low-mass stars are treated as non-rotating in every MIST model given their small rotational velocities \citep{2019ApJ...887..199G}.

\subsection{Mathematical fitting model}
The accurate \emph{Gaia} parallax measurements and precise photometry allowed us to fit isochrones to the `measured' absolute magnitudes using the distance modulus of the individual stars.
Using the prior knowledge from the available spectroscopy discussed in Sect.\,2.1, we restricted the isochrone fitting to solar metallicity. All uncertainties from the photometry were propagated during the model evaluation.

We used a Markov Chain Monte Carlo (MCMC) approach in a Bayesian setting (following e.g. \citealt{2005A&A...436..127J, 2019AJ....158..173A}) and estimated the posterior probability for the cluster age from its single star population, because
binaries can distort the fitting as they do not necessarily occur on the cluster sequence. Due to a known mismatch between the colours of the isochrones and observed colours for low-mass stars (see
\citealt{2016ApJ...823..102C} and references therein), we limited the fitting to $G\leq13$ ($M_G\leq 5.5$, corresponding to masses $\gtrsim 0.9$\,M$_\sun$).

The posterior distributions for the age $t$ and extinction $A_V$ were estimated according to Bayes' theorem:
\begin{equation}
    p(t, A_V | {\bf M}(\varpi)) = \int p({\bf M}(\varpi) | t, A_V) p(t) p(A_V) \mathrm{d}t \mathrm{d}A_V
\end{equation}
with $\bf M$ the combined absolute magnitude measurements of all considered cluster members. The log-likelihood function was given by
\begin{equation}
\begin{split}
    \mathcal{L}_\mathrm{cluster}&=\log p({\bf M}(\varpi) | t, A_V) \\&= -0.5  \sum_i w_i \left[ \left(({\bf M}_i - {\bf I}(M_G, t, A_V))^2 / \sigma_M\right)\right],
\end{split}
\end{equation}
where $\bf I$ contained the isochronal magnitudes given the input parameters and $\sigma_M$ the uncertainties associated with the observations $\bf M$. Each star was weighted with $w_i$ depending on its proximity to the turn-off (in our case the brightest star in the cluster). Stars close to the turn-off were given the highest weights in the fit.

For each observed $M_G$, we interpolated the isochrone linearly and evaluated its $M_\mathrm{BP}$ and $M_\mathrm{RP}$ magnitudes. These values were reddened according to the probed $A_V$ by using the median extinction coefficients from \cite{2018MNRAS.479L.102C}\footnote{These values are strictly speaking only valid up to 7000\,K, yet at the low extinction the effect of over or underestimating the extinction coefficient can be neglected.}.
The prior, $p(t)p(A_V)$, was chosen to be flat and very broad in age ($7 \leq \log t \leq 10.0$) and extinction ($0\,\mathrm{mag} \leq A_V \leq 1\,\mathrm{mag}$). Both values were basically unbound to probe the complete parameter space.
The MCMC calculations were carried out using the open-source \texttt{Python} package \texttt{emcee} \citep{2013PASP..125..306F}. We applied this procedure to each set of isochrones with different rotational velocities.

\subsection{Isochronal age}
Given the absence of evolved cluster members, the age cannot be tightly constrained from the isochrones and spans a wide range independent of the chosen stellar rotation rate. However, the best age (i.e., the age corresponding to the maximum likelihood of the posterior) is strongly correlated with the chosen rotation rate and we find younger ages for faster rotating models (c.f. Fig~\ref{fig:agedistrib} in the Appendix for the age posterior distributions). This effect is not unexpected because UBC\,1 does not host a main-sequence turn-off and colour shifts due to rotation have to be compensated by age differences (and additional extinction to some extent).

\begin{figure*}
    \includegraphics[width=\textwidth]{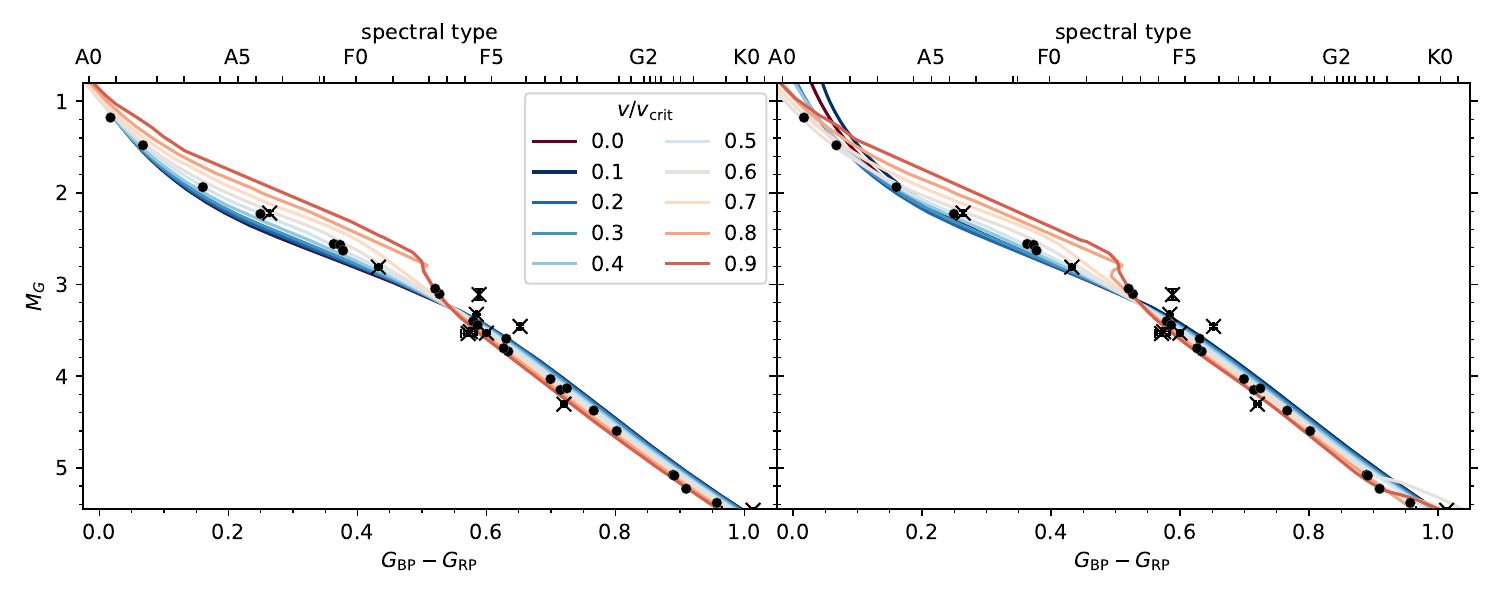}
    \caption{Comparison of best-fitting isochrones with different rotation rates in a colour-absolute magnitude diagram of the upper main sequence stars in UBC\,1. Isochrones shown in the \emph{left} panel use the median values of all samples of a given rotation rate for their parameters (age and extinction). In the \emph{right} panel, we show isochrones based on the maximum likelihood parameters. Crosses indicate binaries and were omitted during the model evaluation. In both cases the $v/v_\mathrm{crit}=0.5$ model describes the data best. The parameters for all isochrones in this figure can be found in Table~\ref{tab:isopara}.}
    \label{fig:rotisocomp}
\end{figure*}

Considering all rotating models would lead to a large age uncertainty of 1\,Gyr in the range $\log t = 7.6 - 8.6$. While the maximum of the posterior distribution changes appreciably depending on the rotation rate, the median age is in the range $\log t = 7.9 - 8.2$.

In Fig.~\ref{fig:rotisocomp}, we show isochrones for the different rotating models based on the median posterior value (left panel) and on the maximum likelihood of the parameters (right panel). It is immediately visible that some isochrones do not match the observations even for their best-fitting parameters (see Table~\ref{tab:isopara} for their values).

In particular, the faster rotating models (redder shades in Fig.~\ref{fig:rotisocomp}) are not able to describe the observations as they overestimate the brightness for many stars. Hence, they are not considered further. Similarly, the slowly rotating models (dark blue shades) have troubles to match the observations. Here, best-fitting values do not match the highest mass stars and provide too old solutions. For these models the joint fit with the cool stars provides additional conditions. Since low-mass stars are considered non-rotating, independent of the higher-mass stars' rotation rate, extinction is the constraining factor for their isochrone position. The extinction added to the slowest rotating models is the cause of the spread observed among the lower-mass stars.

\begin{figure}
    \includegraphics[width=\columnwidth]{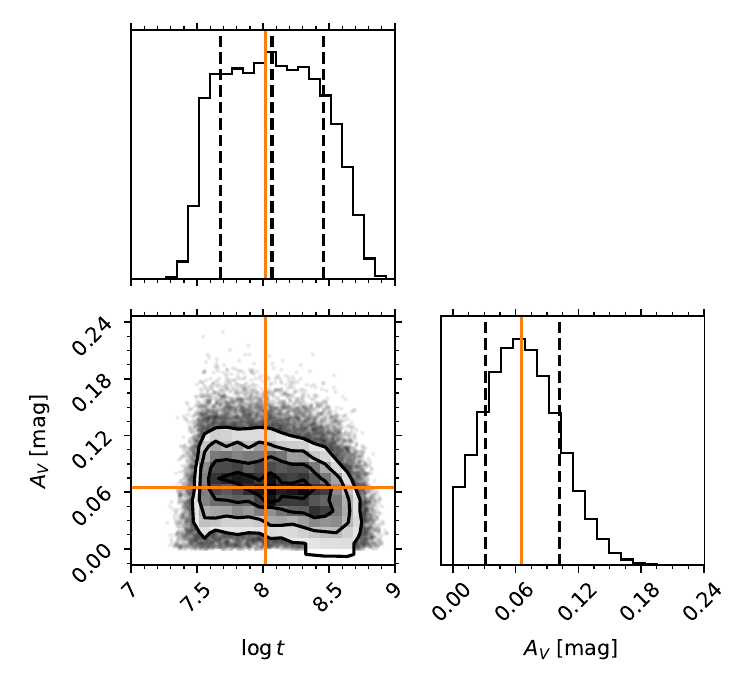}
    \caption{Corner plot of the posterior distributions from the MCMC isochrone fitting of UBC\,1 with the rotating MIST models at $v/v_\mathrm{crit}=0.5$. The orange lines indicates the maximum likelihood values and the dashed marks show the 16th, 50th (median), and 86th percentiles as an uncertainty estimate. For the extinction the median and the maximum likelihood coincide.}
    \label{fig:corner_CMD}
\end{figure}

Based on both panels of Fig.~\ref{fig:rotisocomp}, we find the isochrone with $v/v_\mathrm{crit}=0.5$ to represents the measurements best and to pass through most of the data points. For these models, the maximum likelihood estimator of the age is closest to the median age distribution, which itself is nearly symmetric for this rotation rate (Fig.~\ref{fig:corner_CMD}) while it is heavily skewed for other rotation rates (Fig.~\ref{fig:agedistrib}). Hence, we can safely assume that the selected model and its uncertainty encompasses most of the data.

Reliable rotational velocities ($v\sin i$) are available in the literature for only two stars \citep[APOGEE,][]{2020AJ....160..120J}. These two stars are slow rotators with $v\sin i/v_\mathrm{crit}\approx0.1$. The lack of observational data (from surveys and dedicated studies) for other cluster members does not allow us to draw conclusions on the overall rotational distribution.

The posterior distributions of the most physical model with $v/v_\mathrm{vcrit}=0.5$ is shown in Fig.~\ref{fig:corner_CMD} and leads to the maximum likelihood age $\log(t)=8.1\pm0.4$ with an associated extinction of $A_V=0.065\pm 0.035$\,mag (equivalent to $E(B-V)=0.021\pm0.01$\,mag, assuming $R(V)=3.1$). The uncertainty levels are given by the 16th and 84th percentile of the posterior distribution.

The interstellar extinction value is in good agreement with the rough estimate from \emph{Gaia}. We made several tests to ensure that neither the age nor the extinction depend on the initial value of the MCMC chain or on the prior.

\begin{figure}
    \includegraphics[width=\columnwidth]{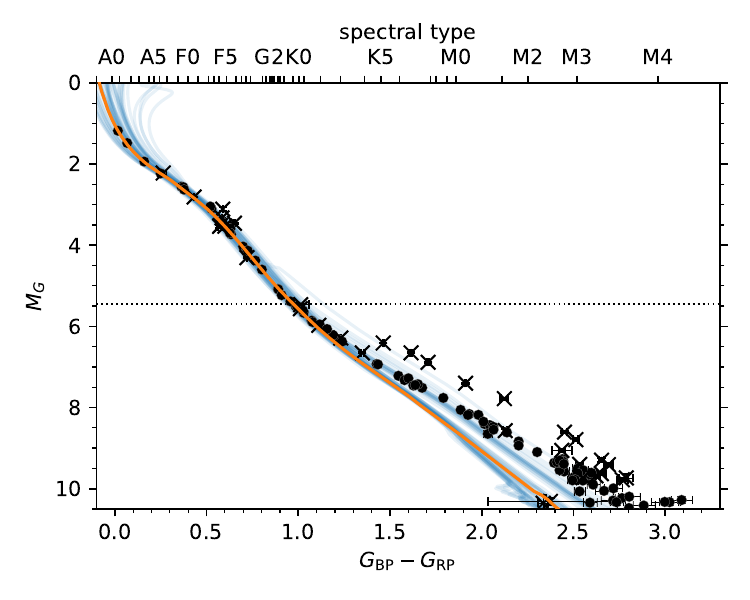}
    \caption{Colour-absolute magnitude diagram of UBC\,1 with a selection of reddened rotating MIST isochrones ($v/v_\mathrm{crit}=0.5$) in blue. The best-fitting isochrone is given in orange and corresponds to the values indicated in Fig.~\ref{fig:corner_CMD}. The dotted horizontal line shows the limit used in the fit. The uncertainties are typically within the symbol size. Crosses indicate potential binaries that were omitted from the fit.}
    \label{fig:CMD_fit}
\end{figure}

Finally, we assess the best-fitting model not only based on the quantitative posterior distributions but also on how it visually fits the data to get a sense of the parameter range and to assess the physical implications of the fit.
In fact, the age of open clusters has often been estimated by visual comparison to isochrones, rather than rigorous mathematical modelling. In Fig.~\ref{fig:CMD_fit}, we show the colour-absolute magnitude diagram of UBC\,1 and over-plot a randomly drawn sample from the posterior distribution, as well as the best fitting model. This re-affirms that the rotating isochrone with the chosen parameters describes the stars in the fitted range very well. Among the randomly drawn models, some stray far from the actual data near the turn-off, providing visual confirmation for the age near 125\,Myr.

To summarize, we find UBC\,1 to be $\log(t)=8.1\pm0.4$ (125\,Myr) old. The reddening towards the open cluster is small with $E(B-V)=0.021\pm0.01$\,mag. We show that it is important to include rotation in the isochrones for CMD fitting based on intermediate-mass stars, even in the case of open clusters without an extended main-sequence turn-off.

\section{TESS observations, reduction, and cluster pulsators}
\label{sec:obs}
\subsection{Observations and data reduction}
\begin{figure}
    \includegraphics[width=\columnwidth]{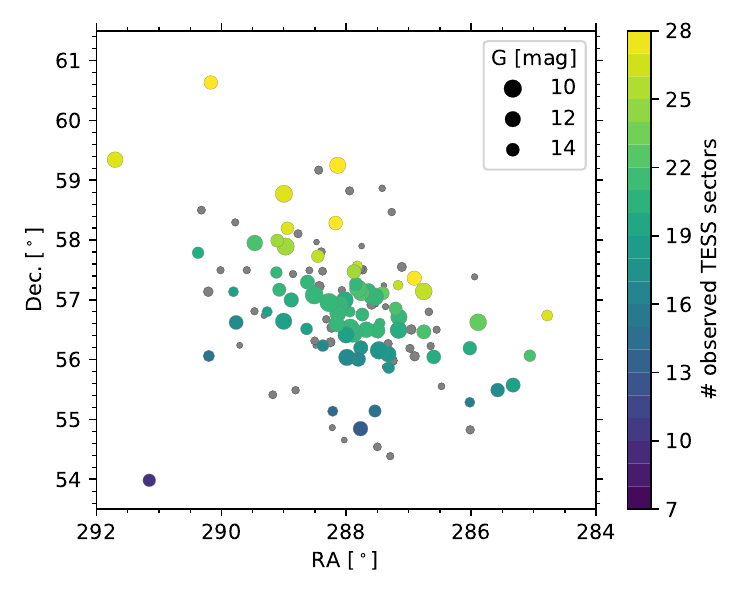}
    \caption{Sky positions of the open cluster members. The colour coding shows the number of available TESS sectors for stars with $G\leq16$\,mag and the size is proportional to their brightness. Fainter members are shown with small grey symbols.}
    \label{fig:sky}
\end{figure}

We chose UBC\,1 as our target because it is situated near the edge of the TESS Northern Continuous Viewing Zone (CVZ-N), delivering uninterrupted time-series photometry up to 352\,d. As seen in Fig.~\ref{fig:sky} not all stars are in the CVZ-N and the coverage fraction is varying. Our targets could potentially be observed in sectors $14-26$ of the primary mission and sectors 40, 41, and $47-60$ of the extended missions (28 sectors in total).

To download and reduce the TESS data, we employed the asteroseismic reduction pipeline developed by \cite{2022A&A...662A..82G}. In short, we gathered all available TESS data for the intermediate-mass cluster members from the Mikulski Archive for Space Telescopes (MAST) using the \texttt{TESScut} API \citep{2019ascl.soft05007B}. We chose a cut-out size of 25\,px x 25\,px, allowing for the identification of neighbouring, contaminating stars and a sufficient background area for subtraction.

For each target, we extracted the flux using custom apertures in the \texttt{Python} package \texttt{Lightkurve} \citep{2018ascl.soft12013L}. The apertures were constructed with the threshold method in \texttt{Lightkurve} (threshold $4.5\sigma$). In order to avoid light of the neighbouring stars in our aperture, we identified relevant neighbours ($\Delta m_T \le 4.5$\,mag) in the cut-out region. We fitted a background planar model to the image and modelled each source as a Gaussian at the background position. The true TESS point spread function is much more complicated but for our purpose a Gaussian estimate is sufficient. With the models for the background and neighbours, we were in a position to estimate the flux contribution of all stars to the pixels within our mask. Pixels in which the contamination fraction is larger than $10^{-4}$ were removed from the aperture mask. Other pixels were kept to avoid too small masks even though it led to irregular mask shapes.

As an additional check for contamination, we selected all \emph{Gaia} sources with $G<16$ within 2\arcmin{} around each A or F-type cluster member and calculated their absolute magnitude. Stars with close neighbours that could lead to potential flux contamination were selected and we checked whether the contaminating star is of spectral type F or earlier (based on their absolute magnitude).
We found only one instance in which this is the case. However, this star is a \emph{Gaia}-resolved equal-mass binary cluster member (UBC1-23 and UBC1-24).

After the extraction, we de-trended each light curve sector-by-sector with a background subtraction and principal component analysis (up to seven components). In the asteroseismic analysis, we are interested in relatively short-term variability, hence removing all long-term trends (longer than one sector) in the light curves
is appropriate. It also allowed us to stitch together all observed TESS sectors into a single light curve. Starting with the extended mission, the cadence of the TESS observations was shortened from 30\,min to 10\,min and with the second extended mission to 200\,s. To create a light curve with a common cadence, we binned all observations of the later sectors to a 30\,min cadence. Due to the location at the edge of the CVZ-N, most star were not observed in all sectors within a cycle, imposing complicated window functions. Yet, the great majority of our stars were completely covered by Sectors 56 through 60, hence we can use these sectors with a shorter cadence to distinguish between effects of the window function and true signals in the frequency analysis.

Our membership list contains eleven stars that could potentially be intermediate-mass pulsators (with spectral types F3 or earlier, $(G_\mathrm{BP}-G_\mathrm{RP})_0<0.5$). We extracted ten light curves. The remaining star (UBC1-73) was too close to a neighbouring star to exclude contamination.

\subsection{Population of pulsators in UBC\,1}
\cite{2022A&A...666A.142S} searched the TESS data of the whole CVZ-N and provide a list of periodically variable A and F stars, including a classification. Three hybrid $\gamma$\,Dor-$\delta$\,Sct pulsators and five stars classified as generally variable are in our membership list. In addition, \cite{2022A&A...666A.142S} also classify UBC1-3 as a $\gamma$\,Dor pulsator. Based on its position in the CMD this star is a K\,dwarf.  We find that rotational modulation of two nearby periods in the periodogram of this star (maybe caused by latitudinal differential rotation) generates a pattern that resembles a period spacing pattern at first look. Three stars were not classified but are included in \cite{2022A&A...666A.142S}. Their list of known pulsators serves as our initial list which we aim to expand and analyse in detail.

\begin{figure}
    \includegraphics[width=\columnwidth]{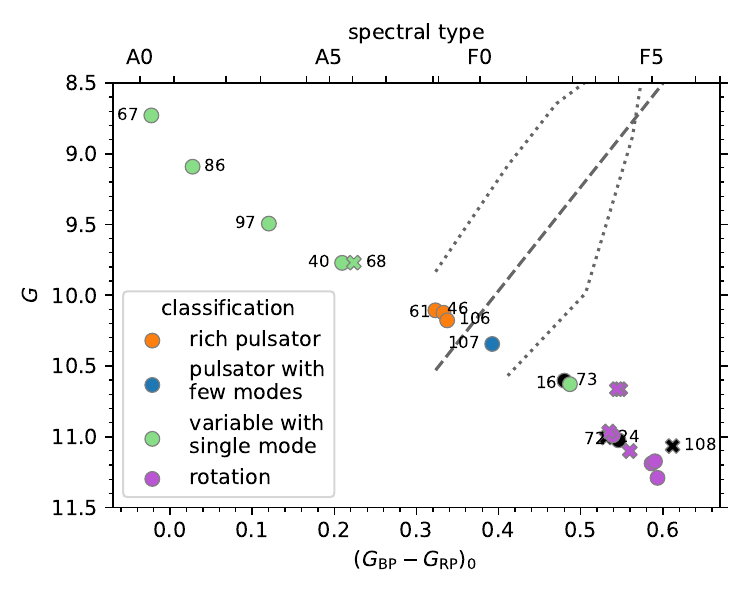}
    \caption{Colour-magnitude diagram of UBC\,1 with the intermediate-mass members colour-coded by their number of detected signals. The three rich pulsators are shown in orange. In blue, we show the pulsator with only a handful of independent frequencies in their periodogram, while stars in green have either a single frequency or noisy Fourier spectra and can therefore not be identified as pulsators unambiguously. For stars with $(G_\mathrm{BP}-G_\mathrm{RP})_0>0.5$, rotational modulation (purple) is the only source of variability. Black symbols indicate stars without a light curve. Stars shown as crosses are potential binaries. The dotted lines indicate the approximate edges of the $\gamma$\,Dor instability strip from \cite{2005A&A...435..927D} and the dashed line denotes the red edge of the empirical $\delta$\,Sct instability strip from \cite{2019MNRAS.485.2380M}. Details on individual stars (labelled with their ID number) can be found in the text.
    }
    \label{fig:CMDpul}
\end{figure}

Our initial classification of pulsators is carried out with Lomb-Scargle periodograms \citep{1976Ap&SS..39..447L, 1982ApJ...263..835S} generated by the asteroseismic pipeline \citep{2022A&A...662A..82G}. We find the majority of stars earlier than F5 ($(G_\mathrm{BP}-G_\mathrm{RP})_0<0.6$) to show at least some level of periodic variability as they reveal peaks in their periodogram, which occur clearly above the noise level. These stars are highlighted in Fig.~\ref{fig:CMDpul} and colour-coded by their perceived richness of the periodogram from an asteroseismic perspective.

Three stars (UBC1-46, UBC1-61, UBC1-106) show very clear pulsation signals. These three stars are already classified as hybrid pulsators by \cite{2022A&A...666A.142S} and are discussed in detail below. The star UBC1-107 shows only a few periodic components in the frequency domain. Yet, we interpret these as pulsation frequencies rather than rotational signals as it concerns values in the range of a few cycles per day. This member is described as variable by \cite{2022A&A...666A.142S}. In Fig.~\ref{fig:CMDpul}, we also mark the theoretical instability strip for g-mode pulsations from \cite{2005A&A...435..927D} and the observed red edge of the p-mode strip deduced by \cite{2019MNRAS.485.2380M}\footnote{We note that transforming instability strip edges from $\log L/L_\sun$ and $T_\mathrm{eff}$ to \emph{Gaia} magnitudes involves empirical transformations which might introduce biases. They are only meant as a guidance.}. These four stars are located in the $\gamma$\,Dor instability strip and are thus expected to pulsate in g modes. Moreover, the three richest pulsators show both g- and p-mode pulsations and are located in the narrow area that is shared by both instability regions.

The remaining variable stars (UBC1-16, UBC1-40, UBC1-67, UBC1-68, UBC1-86, UBC1-97) show maximally one or two significant high-frequency components. Among them UBC1-16, UBC1-67, UBC1-73, and UBC1-86 are described as variables by \cite{2022A&A...666A.142S}. All of these stars are more massive than typical $\gamma$\,Dor stars and occur above the instability region while their variability has frequencies in $1<f<10$\,d$^{-1}$ only. We do not find high-frequency p-modes typical of young $\delta$\,Sct-type pulsators for these stars, despite their position within the classical instability strip.

In the following, we give more details on the variable stars focussing on the three multi-periodic g-mode pulsators.
We show the amplitude spectra in the range $[0,65]$\,d$^{-1}$ as we found that none of the stars reveal
periodic variability beyond this frequency.

\subsubsection{UBC1-46}
\begin{figure}
    \includegraphics[width=\columnwidth]{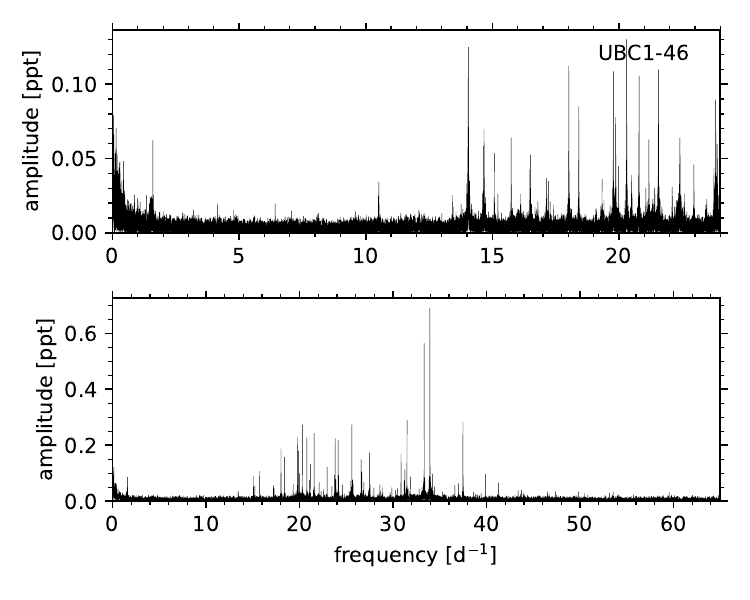}
    \caption{Amplitude spectra of UBC1-46. \emph{Top}: Periodogram of the full data set of 21 TESS sectors with a cadence of 30\,min.
    \emph{Bottom:} Periodogram including only data from Sectors $56-60$ with a 200\,s cadence.}
    \label{fig:pg46}
\end{figure}

UBC1-46 (TIC 406930461) is a genuine $\delta$\,Sct star with some low-amplitude frequencies in the $\gamma$\,Dor g-mode regime. The frequencies do not lead to mode identification, hampering asteroseismic inferences. In particular, we cannot identify a period spacing pattern from the low-amplitude g modes. Both periodograms in Fig.~\ref{fig:pg46} also show a single peak at 1.6\,d$^{-1}$, which might be the star's surface rotation frequency or a (sub-)multiple thereof.

\subsubsection{UBC1-61}
\begin{figure}
    \includegraphics[width=\columnwidth]{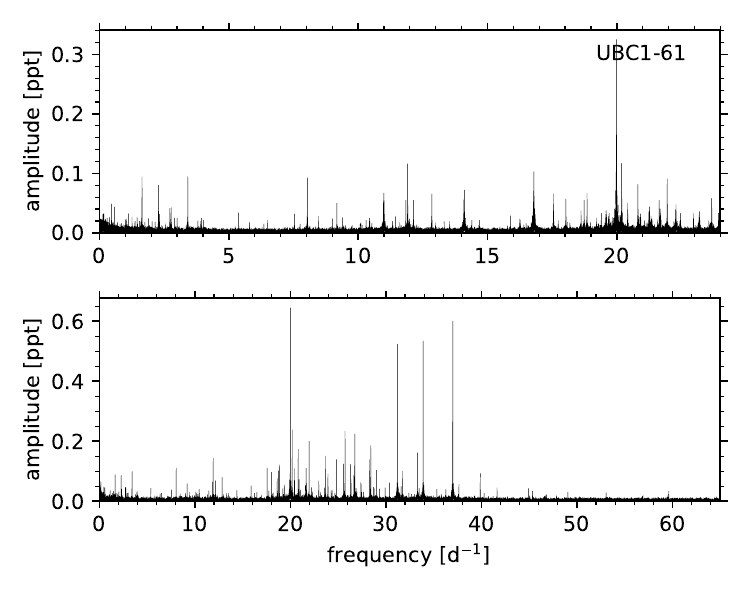}
    \caption{Frequency spectra of UBC1-61 based on 22 sectors.
    See Fig.~\ref{fig:pg46} for more information.}
    \label{fig:pg61}
\end{figure}

Figure~\ref{fig:pg61} shows the amplitude spectra of the richest hybrid pulsator in the cluster. This star (UBC1-61, TIC 416402521) does not only show a wealth of higher-frequency pulsations but also many peaks from the low-frequency domain of g-mode pulsations all the way to the p-mode regime.
Despite its abundance in pulsation frequencies, we are not able to identify a period spacing pattern. We suspect that multiple overlapping patterns with missing peaks occur. We also do not find obvious combination frequencies neither in the g- nor in the p-mode regime.

\subsubsection{UBC1-106}
\begin{figure}
    \includegraphics[width=\columnwidth]{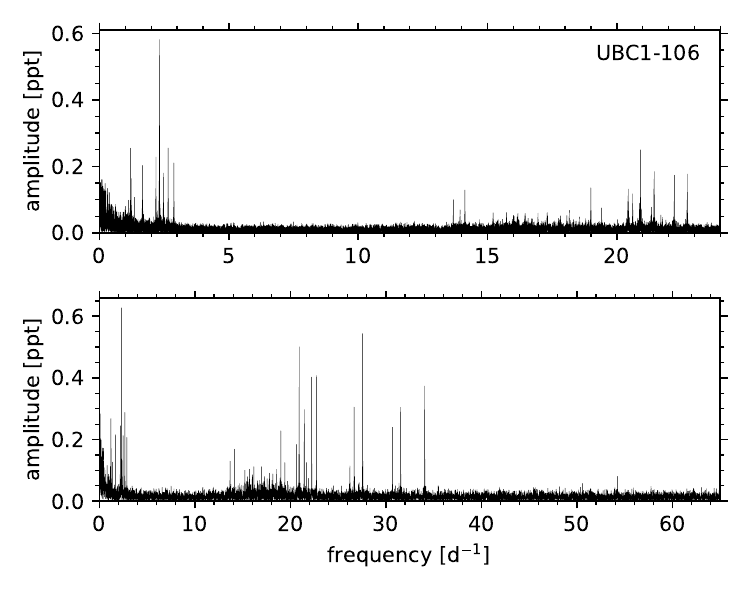}
    \caption{Frequency spectrum of UBC1-106 based on 25 sectors. This star shows a clear period spacing pattern. See Fig.~\ref{fig:pg46} for more information.}
    \label{fig:pg106}
\end{figure}

Although UBC1-106 (TIC 421334449) is very similar to both stars discussed above in terms of its position in the CMD and hence mass, its periodogram is very different. This star is also a hybrid pulsator but only a few prominent peaks in the g-mode frequency regime occur (Fig.~\ref{fig:pg106}). However, these exhibit a clear, comb-like signature of a period spacing pattern. Unlike UBC1-46 and UBC-61, the highest amplitude peak in the amplitude spectrum of UBC1-106 occurs among the lower frequencies. We identify and interpret the period spacing pattern in Sect.~\ref{sec:asteroage}.

\subsubsection{Other variable cluster members}
Among the remaining variables, we are not able to identify stars with clear pulsation signals in their periodograms. These cluster members are periodic variables and exhibit one or multiple periodic components in the g-mode regime. However, their individual detected modes are currently not suited for asteroseismic inferences as long as they cannot be identified in terms of degree and azimuthal order. High-resolution spectroscopy may remedy this and still lead to asteroseismic modelling \citep[cf., cases of MOST and CoRoT B-type pulsators, such as in][]{Handler2009,Aerts2011,2018A&A...616A.148B,Aerts2019-CoRoT}. Here, we only discuss the targets and show their periodograms in Appendix~\ref{app:othervariables}.

The two most interesting pulsating cluster members of this class are UBC1-67 (TIC 416403139) and UBC1-107 (TIC 421335009). The former might be a ``hump-and-spike'' star \citep{Henriksen2023}, in which the observed feature could originate from Rossby modes \citep{2018MNRAS.474.2774S}. If this is the case, we estimate the rotation frequency of UBC1-67 to be $f_\mathrm{rot}\approx 1.6$\,d$^{-1}$. However, the frequency peak could also be a solitary pulsation or surface rotation frequency. We note that UBC1-67 is the brightest and hence highest mass star in the open cluster.

For UBC1-107, we find several frequencies below $f=3$\,d$^{-1}$, which could be part of a period spacing pattern. However, they are very close to the noise level or even insignificant, hence we are not able to construct a reliable period spacing pattern as there would be missing modes of consecutive radial order.

In case these features in frequency space originate from pulsations, the two stars might be promising targets to revisit once additional TESS observations are available. A longer time baseline would provide higher frequency resolution of nearby pulsation peaks and might elevate additional pulsation frequencies above the noise level.

Further, we note that UBC1-40 (TIC 406925885) shows low level stochastic variability with $f\leq 2$\,d$^{-1}$. The frequency spectrum of UBC1-86 (TIC 243275988) has an isolated frequency around $f\approx5.1$\,d$^{-1}$, while its spectrum is otherwise featureless for $f\gtrsim1$\,d$^{-1}$.

\section{Asteroseismic age-dating of UBC\,1}
\label{sec:asteroage}

The main aim of this work is to provide an asteroseismic age constraint for UBC\,1 based on identified g modes in its cluster members and to confront this asteroseismic age with the isochronal and gyrochronal ages. In order to derive an asteroseismic age, we attempt to estimate the buoyancy travel time ($\Pi_0$), which is an internal structure quantity characterizing the size of the mode cavity where the detected g modes propagate. $\Pi_0$ can be used as an age-indicator for intermediate-mass stars because it probes the near-core g-mode cavity that changes during the main sequence evolution. Generally speaking, the shrinking convective core leaves behind a chemical gradient which increases the stability against convection \citep{1947ApJ...105..305L} and hence the buoyancy frequency (Brunt-V\"ais\"al\"a frequency) increases. Thus, the inversely related buoyancy travel time decreases with age.

Observationally, the buoyancy travel time is delivered by period spacing patterns of g modes \citep{2016A&A...593A.120V,Ouazzani2017}.
In order to infer $\Pi_0$ from observations, an estimate of the near-core rotation frequency, $f_{\rm rot}$, is needed because it is determined by the slope of the period spacing pattern and slightly correlated with $\Pi_0$ \citep{2013MNRAS.429.2500B}. To estimate $f_{\rm rot}$, the identification of the spherical wave numbers $(l,m,n)$ of the modes involved in the period spacing pattern is required. We use the methodology initially developed by \citet{2016A&A...593A.120V} and improved by \citet{2018A&A...618A..24V} to achieve mode identification and estimation of $\Pi_0$ and $f_{\rm rot}$. As a first step in the application of this method, we hunt for period spacing patterns of g modes for the cluster pulsators.

\subsection{Pre-whitening}
We analysed the three pulsators with clear g modes in detail to obtain asteroseismic parameter estimates. For that purpose, we extracted all their significant frequencies from the periodograms through iterative pre-whitening. For this work, we used the frequency analysis routines developed by \cite{2021A&A...655A..59V}. We used the option to deduce all frequencies with a signal-to-noise ratio (S/N) threshold $\mathrm{S/N}\geq3$. The S/N was calculated in a window of 1\,d$^{-1}$ around the extracted frequency. We chose this low cut-off S/N value as stopping criterion to extract the significant frequencies compared to the often cited 5.6 for space photometry \citep{baran2015,burssens2019}
because we possess light curves with difference cadences that allows us to cross-check whether peaks are significant.

With the lists of pre-whitened mode periods, we proceeded to search for g-mode period spacing patterns. To facilitate the identification, we used the open-source \texttt{Python} package \texttt{FLOSSY}\footnote{\url{https://github.com/IvS-KULeuven/FLOSSY}} \citep{2022A&A...662A..82G}. It allows to visually fit the observed periods by displaying a period spacing plot and an \'echelle diagram. When a matching pattern is found, \texttt{FLOSSY} calculates the $\chi^2$ value to check whether the selected pattern is locally the optimal solution. With this visual approach, FLOSSY allows the user to easily select which mode periods belong to the pattern and investigate whether additional modes with frequencies below the selection threshold might be present in the periodogram.

\subsection{Period spacing pattern for UBC1-106}
\begin{figure}
    \includegraphics[width=\columnwidth]{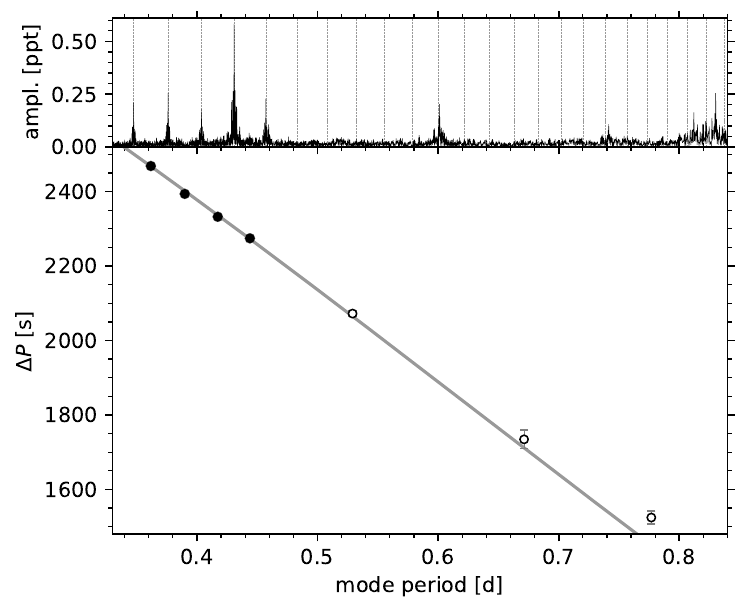}
    \caption{Period spacing pattern of UBC1-106. The \emph{top} panel shows the periodogram with the modelled pattern as lines. \emph{Bottom:} Fit (grey line) to the period spacing pattern including the five prograde dipole modes of consecutive radial orders $n=10, 11, 12, 13, 14$ (filled circles). The open circles are additional modes with significant frequency that might belong to the pattern as well, but this is uncertain from the current TESS data. The markers are placed at the mean position between the detected signals and hence fall in between the modes marked in the upper panel.
    }
    \label{fig:psp106}
\end{figure}

As outlined above, the only successfully identified period spacing pattern belongs to UBC1-106. Using \texttt{FLOSSY}, we find a
pattern of five modes with consecutive radial order. The period spacing pattern of these five dominant modes is very regular, as illustrated in Fig.~\ref{fig:psp106}.

We used the smooth pattern to perform mode identification and estimate the buoyancy travel time ($\Pi_0$) and near-core rotation frequency $f_\mathrm{rot}$ from the formalism derived in \cite{2016A&A...593A.120V} and updated in \cite{2018A&A...618A..24V}. This procedure was already successfully applied to TESS g-mode field pulsators by \citet{Garcia2022b}. For UBC1-106 this
led to the identification of five prograde dipole modes of consecutive radial order $n\in[10, 14]$.
The corresponding buoyancy travel time was estimated to be $\Pi_0=4549\pm24$\,s and we deduced a relatively low near-core rotation frequency of $f_\mathrm{rot} = 0.544\pm0.009$\,d$^{-1}$. Assuming rigid rotation, the near-core rotation rate corresponds to $v_\mathrm{surf} = 46$\,km\,s$^{-1}$ which is in agreement with the spectroscopically measured surface rotation $v\sin i =22.6$\,km\,s$^{-1}$ from APOGEE \citep{2020AJ....160..120J} and points to an inclination angle of about 30\degr. Dips due to the occurrence of mode trapping caused by a chemical gradient are typically observed in evolved stars \citep{2008MNRAS.386.1487M, Kurtz2014, 2019MNRAS.485.3248M}. The smooth pattern thus hints towards a young pulsator, in agreement with the isochrone fitting.

In addition to the smooth pattern centred around mode periods of 0.4\,d, we find three additional modes with longer periods that could
belong to the pattern as well. These mode periods are seemingly separated from the pattern. Should they belong to it, they must have higher and non-consecutive radial orders. Without the certainty of these modes belonging to a long pattern, we can neither rule out nor confirm that the observed modes belong to the same series as the modes with periods around 0.4\,d. Hence, we continue with the asteroseismic parameters determined from the five consecutive radial order modes.

\subsection{Asteroseismic age of UBC1-106}
In order to estimate an asteroseismic age for UBC1-106, additional astrophysical observables, aside from $\Pi_0$, are beneficial, provided that they are of high precision \citep[see e.g.][]{2019MNRAS.485.3248M}. We first considered the effective temperature ($T_\mathrm{eff}$) and surface gravity ($\log g$) from \emph{Gaia} DR3 \citep{2023A&A...674A..26C}. However, these spectroscopic values have unrealistically small errors and their values in the supplementary astrophysical parameter tables also disagree.
For this reason, we resorted to the stellar luminosity as an additional independent observable from \emph{Gaia} DR3. Given the absolute magnitude of UBC1-106 and the bolometric correction \citep{2023A&A...674A..26C}, we obtained $\log L/L_\sun=0.82\pm0.02$. While the bolometric correction is model dependent and also depends on the effective temperature, we checked that it is insensitive to $T_\mathrm{eff}$ in the considered temperature range (see also Fig.\,8 in \citealt{2018A&A...616A...8A}). Our result for the luminosity was therefore essentially unaffected by the poor uncertainty estimate for the effective temperature from Gaia DR3 \citep{2023A&A...674A..26C}.

\begin{figure*}
    \includegraphics[width=0.5\textwidth]{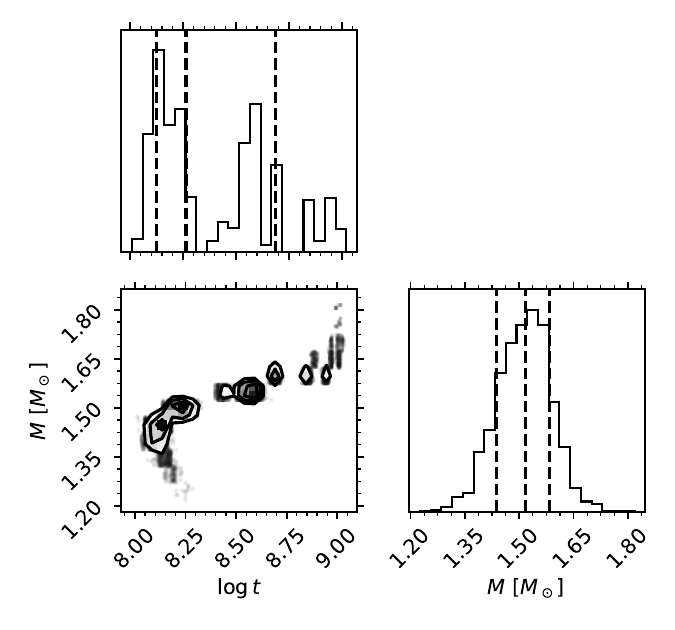}
    \includegraphics[width=0.5\textwidth]{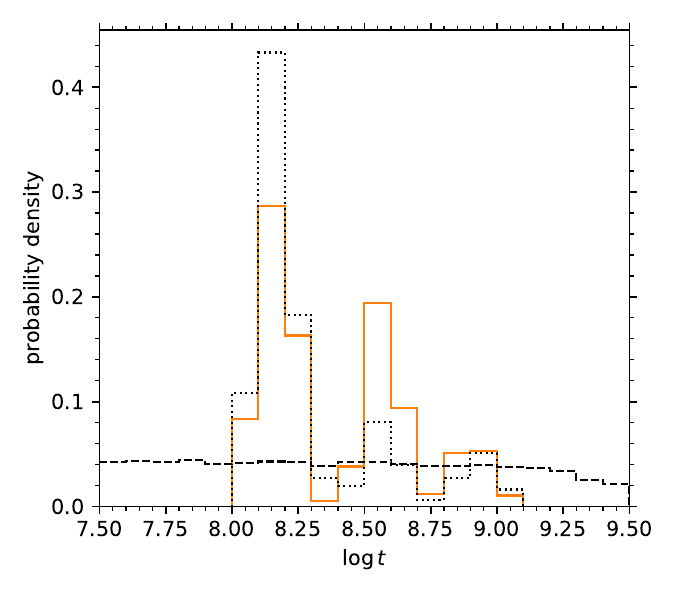}
    \caption{Posterior distributions resulting from the MCMC grid search. \emph{Left:} corner plot for the grid search based on $\Pi_0$ and $\log L/L_\sun$ as observables. Both the age and the mass of UBC1-106 are well constrained despite the correlation between the two parameters. The dashed lines indicate the 16th, 50th (median), and 66 percentile of the distribution.
        \emph{Right:} Posterior age distribution based on \emph{Gaia} observables ($\log L/L_\sun$, $T_\mathrm{eff}$, and $\log g$) alone (black dashed), only $\Pi_0$ (black, dotted), and $\Pi_0$ and $\log L/L_\sun$ (solid, orange, same as top in \emph{left} panel, the differences are due to a finer binning). Without the asteroseismic information, the age of UBC1-106 cannot be constrained.
    }
    \label{fig:corner_astero}
\end{figure*}

To estimate UBC1-106's age from $\Pi_0$ and $\log L/L_\sun$, we developed an MCMC grid search similar to the isochrone fitting in Sect.~\ref{sec:isoage}, but this time relying on the dedicated MESA (\texttt{v7385}) stellar structure and evolution grid of models for $\gamma\,$Dor stars computed by \cite{2016A&A...593A.120V}. It was calculated for stars within the mass range $1.2 \leq M_\star/M_\sun \leq 2$. For each stellar mass various combinations of the initial chemical composition, convective core overshoot values for both an exponentially decaying and a step overshoot description, and diffusive envelope mixing levels are available. We refer the reader to \cite{2016A&A...593A.120V} for details of the input physics omitted here and point out that it is different from the one in the MIST isochrones (c.f. \citealt{2016ApJ...823..102C}). Following the cluster's general properties, we limited the grid search to models with solar metallicity ($Z=0.014$, $X_\mathrm{ini}=0.71$).

For the measured $\Pi_0=4549\pm24$\,s, the application of the grid search delivered the posterior distributions for the stellar mass and age shown in Fig.~\ref{fig:corner_astero}. We find a well defined lower boundary in age. The mean value (coinciding with the maximum likelihood) is $\log t= 8.24^{+0.43}_{-0.14}$ ($t = 175^{+280}_{-50}$\,Myr) for a $1.51\pm0.08$\,M$_\sun$ star with an exponential overshoot $f_\mathrm{ov}=0.0075$ (in agreement with \cite{2019MNRAS.485.3248M, 2019ApJ...876..134C}; the results do not change for the best model with a step overshoot). The envelope mixing is not well constrained, as it is also the case for the even more precise \emph{Kepler} $\gamma\,$Dor asteroseismology \citep{Mombarg2021}. Its value does not affect the mass and age results. Hence, we fixed it at $D_\mathrm{mix}=1$\,cm$^2$\,s$^{-1}$.

The asteroseismic age estimate for UBC1-106 is well in agreement with the isochronal age of the cluster (Sect.~\ref{sec:isoage}). As it is common for asteroseismology of stars with a convective core and overshooting, we find a correlation between mass and age \citep[see][]{2019MNRAS.485.3248M}, explaining the larger upper than lower age uncertainty. Despite the seemingly good agreement, we keep in mind that systematic uncertainties due to the choice of input physics in stellar models occur but are still largely unknown. Detailed asteroseismic analyses of open clusters will eventually help in reducing these systematic uncertainties.

\begin{figure*}
    \includegraphics[width=\textwidth]{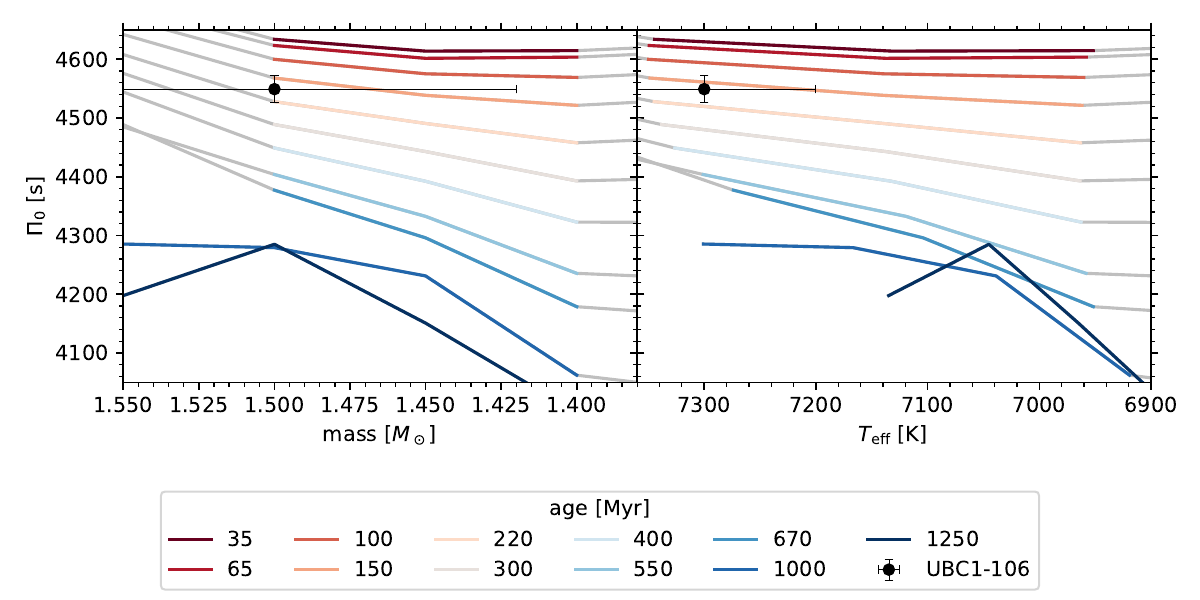}
    \caption{Position of UBC1-106 among isochrones in the plane of $\Pi_0$ against mass (\emph{left} panel) and $T_\mathrm{eff}$ (\emph{right} panel), respectively. The coloured part of each line indicates the mass range in which the isochrone crosses the theoretical $\gamma$\,Dor instability strip of \cite{2005A&A...435..927D}. The clear separation of the isochrones highlights the potential that g-mode asteroseismology holds for age-dating young open clusters.
    }
    \label{fig:Pi0_iso}
\end{figure*}

An observationally motivated approach of determining the age of UBC\,1 is to place the observed properties onto isochrones in the mass-$\Pi_0$ plane. Fig.~\ref{fig:Pi0_iso} shows these isochrones for different ages and constructed from the MESA grid of \cite{2016A&A...593A.120V}.
In subsequent studies, we will populate this diagram with more $\gamma$\,Dor stars in open clusters to empirically calibrate the asteroseismic models and probe systematic uncertainties in the determination of $\Pi_0$. Our observed star has uncertainties that cross the borders of the instability strip of \cite{2005A&A...435..927D}. It is well known that g-mode pulsators can be found beyond that region \citep[see e.g.][]{2023A&A...674A..36G}. The $\gamma\,$Dor stars have two different types of mode excitation mechanisms active in them \citep{2010aste.book.....A} and, moreover, the instability strips are computed for just one choice of input physics, usually ignoring internal rotation. However, $\Pi_0$ carries solely information on the stellar structure and is independent of the g-mode excitation mechanism.

For completeness, we point out that we also ran a grid search with just the \emph{Gaia} observables $\log L/L_\sun$, $T_{\rm eff}$, and $\log\,g$ instead of relying on $\Pi_0$ and $\log L/L_\sun$. For this to give meaningful results, we had to inflate the errors of $T_{\rm eff}$, and $\log\,g$ arbitrarily \citep[see also ][]{2023A&A...674A..36G}. This led to more uncertain and quite broad posterior age and mass distributions (right panel of Fig.~\ref{fig:corner_astero}), which is not surprising given UBC1-106 is a main sequence star. Hence, the inclusion of the asteroseismic information contained in $\Pi_0$ strongly constrains the posterior and allows to establish much more precise stellar parameters. In fact, using only the asteroseismic observable $\Pi_0$ also leads to a well-peaked posterior distribution for the mass and age.

\subsection{Large frequency separation of p-modes}
Similar to the period spacing pattern in the g-mode regime, p-modes in $\delta$\,Sct stars can show a regular frequency spacing in the absence of fast rotation. In that case, mode identification from such patterns is sometimes possible, particularly for young stars \citep{Bedding2020}. If mode degrees can be identified, this pattern may give rise to the so-called large frequency separation, $\Delta \nu$, which is the difference in frequency between modes of the same degree and consecutive radial order. This separation can be used in a similar manner as $\Pi_0$ to constrain the stellar age \citep{2023MNRAS.526.3779M}. The p-modes detected in the three $\delta$\,Sct pulsators of UBC\,1 occur between $\sim{}15$\,d$^{-1}$ and $\sim{}40$\,d$^{-1}$ which is not enough to construct a pattern because potential patterns are too short hindering an unambiguous identification of the modes. Hence, we cannot give an age estimate for UBC\,1 based on the p-mode pulsations.

\section{Gyrochronal age of UBC\,1}
\label{sec:gyroage}
As an open cluster, UBC\,1 contains not only intermediate-mass stars but hosts also a large number of low-mass members. These stars with outer convection zones shed angular momentum through their magnetized stellar wind \citep{1958ApJ...128..664P, 1967ApJ...148..217W, 1972ApJ...171..565S}, which effectively spins them down on the (pre-)main sequence. The interconnection between rotation, the stellar magnetic field and the field creating dynamo leads to a feedback mechanism that reduces the rotation periods of low-mass stars to a function of mass and age \citep{2003ApJ...586..464B, 2010ApJ...722..222B}. (Metallicity is an additional parameter but basically all open cluster accessible to time-series photometry of low-mass stars have a near solar metallicity.) This property makes cool star rotation periods a powerful tool to determine stellar ages and ages of coeval populations in particular \citep[e.g.][]{2019AJ....158...77C,2022AJ....163..275A,2023A&A...674A.152F}.

With our frequency analysis workflow detailed above, we can derive rotation periods for a number of cool star members of UBC\,1. At the anticipated age of this open cluster, we expect most observable stars to rotate with $P_\mathrm{rot}\lesssim10$\,d, which requires no changes to our photometric procedure adopted above. Hence, we applied the same pipeline to all low-mass members with $G<16$\,mag.

Even with TESS space photometry, the rotation periods can be prone to aliasing, in particular with a factor of two. Hence, we manually verified the detected periods in each light curve and made sure that each detected period is a good representation of the rotational variability in the light curve (see Appendix~\ref{app:LCs} for the light curves of the final selection of stars). Stars for which we were not able to confirm that the periodicity found in the Fourier transform corresponds with the behaviour in the time domain were rejected. This leaves us with 17 rotational period measurements (Table~\ref{tab:periods}) among 62 potential low-mass cluster stars
within the adopted brightness limit of 16\,mag. This relatively low yield is a consequence of our rather strict rejection of members whose TESS light curve are potentially contaminated by close-by sources. This criterion has a bigger influence on the aperture masks of fainter cluster stars. Since our aim is not to provide a detailed rotational analysis of the open cluster but rather a confrontation between gyrochronology and asteroseismic age-dating, this modest yield is acceptable.

\begin{table}
    \caption{Rotation periods for cool star members of UBC\,1.}
    \label{tab:periods}
    \begin{tabular}{rrrr}
        \hline
        \hline
        ID & RA & Dec. & $P_\mathrm{rot}$\\
        & ($\deg$) & ($\deg$) & (d)\\
        \hline
        5 & 285.88908 & 56.62504 & 7.14\\
        12 & 286.59960 & 56.04614 & 7.13\\
        15 & 286.75840 & 56.46539 & 6.81\\
        16 & 286.76079 & 57.14138 & 1.53\\
        22 & 287.15771 & 56.71201 & 1.48\\
        32 & 287.32825 & 56.09310 & 2.48\\
        51 & 287.68033 & 56.50093 & 3.97\\
        59 & 287.77272 & 54.84590 & 4.68\\
        60 & 287.80944 & 56.00957 & 5.47\\
        65 & 287.87613 & 57.47255 & 7.57\\
        77 & 288.13583 & 56.77234 & 2.20\\
        78 & 288.13763 & 59.25090 & 2.15\\
        79 & 288.15372 & 56.59593 & 4.06\\
        80 & 288.17008 & 58.28508 & 7.92\\
        99 & 288.62189 & 57.29618 & 5.79\\
        123 & 290.16919 & 60.63498 & 8.22\\
        129 & 291.70177 & 59.34425 & 3.81\\
        \hline
    \end{tabular}
\end{table}

\begin{figure*}
    \includegraphics[width=\textwidth]{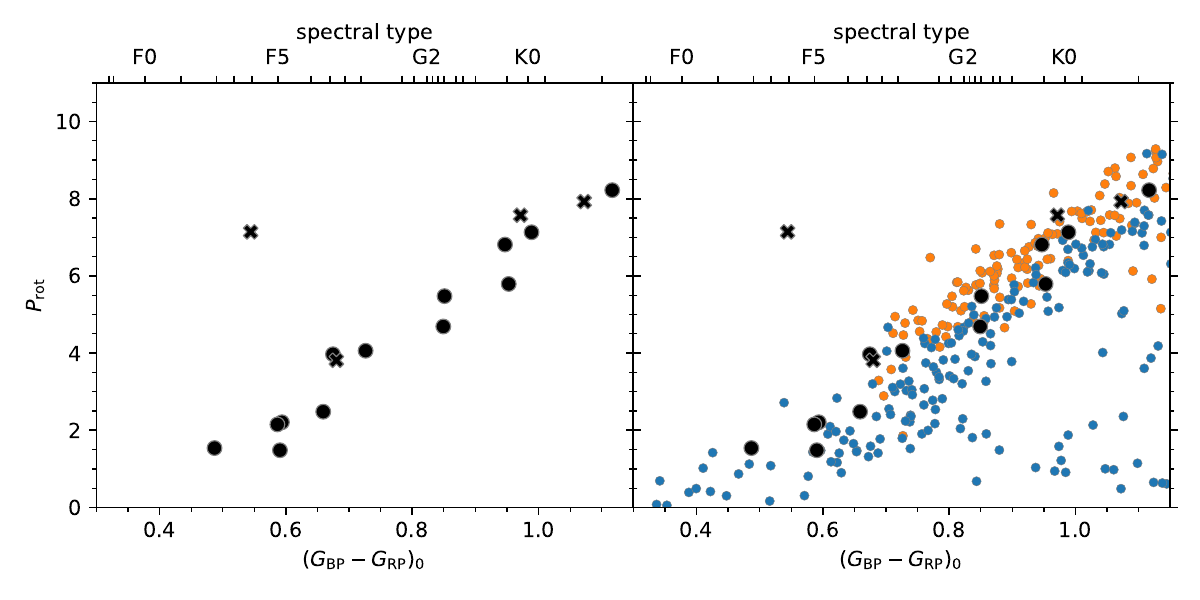}
    \caption{Colour-period diagram of UBC\,1. \emph{Left:} rotation periods of 17 stars in UBC\,1 against \emph{Gaia} $(G_\mathrm{BP}-G_\mathrm{RP})_0$. \emph{Right:} Comparison of the rotators in UBC\,1 (black) with those in NGC\,2516 ($\sim$150\,Myr, blue) and NGC\,3532 ($\sim$300\,Myr, orange). Stars in UBC\,1 generally rotate slower than those in NGC\,2516 while faster than or similar to those of NGC\,3532. Crosses in both panels indicate potential binaries in UBC\,1.
    }
    \label{fig:CPD}
\end{figure*}

Cool star rotation periods are best discussed in a mass-dependent way for which typically an observational proxy of the mass is used. We show the colour-period diagram of the 17 low-mass members with a rotation period estimate in Fig.~\ref{fig:CPD} (left panel) using \emph{Gaia} $(G_\mathrm{BP}-G_\mathrm{RP})_0$. The rotation periods exhibit the expected shape, with fast rotation for the highest-mass stars and slower rotation with decreasing mass.

While the majority of stars follow the expected sequence, one obvious outlier can be found above it. This star (UBC1-5) was identified as a potential binary and we might have picked up the rotational signal from its lower-mass companion or a period connected with the orbit. This scenario would also explain why we observe the rotational modulation only in a few sectors\footnote{The position angle of the TESS field influences the observability of rotation periods with TESS in crowded fields including close binaries.}.

The power of gyrochronology in age determination comes from relative comparisons with populations of known age. Based on the above determined isochronal and asteroseismic ages we selected the two southern open clusters NGC\,2516 \citep[$\sim$150\,Myr,][]{2007MNRAS.377..741I, 2020A&A...641A..51F, 2020ApJ...903...99H, 2021AJ....162..197B} and NGC\,3532 \citep[$\sim$300\,Myr,][]{2021A&A...652A..60F} as comparison clusters. As seen from Fig~\ref{fig:CPD} (right panel), both clusters have low-mass stars with rotation periods bracketing those of the stars in UBC\,1. We find that G and K-type stars rotate slower in UBC\,1, than in NGC\,2516, while similarly to or faster than those in NGC\,3532 within the limits of the comparison due to the scatter in both sequences.
Hence, UBC\,1  has a rotational age determined from its low-mass rotational variables between 150 and 300\,Myr ($t=230\pm70$\,Myr).

\section{Discussion and conclusions}
\label{sec:discuss}
In this work, we provide age constraints from g-mode asteroseismology, gyrochronology, and isochrone fitting for the recently discovered open cluster UBC\,1. Its position near the edge of the TESS Northern continuous viewing zone makes it a prime target to attempt g-mode asteroseismology, which is dependent on long time-series of high-precision photometric observations to resolve individual pulsation frequencies.

Based on a spatial-kinematic filtering of \emph{Gaia} data, we find 132 members in UBC\,1. After establishing the cluster membership, we estimate an isochronal age for UBC\,1. A rotating isochrone with $v/v_\mathrm{crit}=0.5$ describes the \emph{Gaia} data best and we estimate the age of UBC\,1 to be $\log t = 8.1\pm0.4$ ($t = 126^{+190}_{-76}$\,Myr). We also find a very small reddening towards the open cluster with $E(B-V)=0.021\pm0.01$\,mag.

The TESS observations of the intermediate-mass stars in UBC\,1 provide a wealth of variable stars. Three stars are hybrid pulsators located in the small overlap region of the $\gamma$\,Dor and $\delta$\,Sct instability strips. Despite the plethora of observed variability, including many pulsations, we can only identify the pulsation modes of one star thanks to our successful construction of its g-mode period spacing pattern. We use UBC1-106's identified dipole prograde modes to deduce its buoyancy travel time ($\Pi_0$) and near-core rotation
frequency ($f_{\rm rot}$). Asteroseismic modelling of this single cluster member delivers an age of
$\log t= 8.24^{+0.37}_{-0.12}$ ($t = 175^{+232}_{-43}$\,Myr). We stress that we can constrain the asteroseismic age of UBC\,1 with only the measured  $\Pi_0$ from TESS and the {\it Gaia\/} luminosity of a single cluster member without the need for precise spectroscopic information.

We use the TESS light curves of the low-mass members of UBC\,1 to estimate a rotational age for the cluster. Through a comparison of the rotation period distribution with other open clusters (NGC\,2516 and NGC\,3532), we find the low-mass members of UBC\,1 to be $230\pm70$\,Myr old ($\log t = 8.35^{+0.16}_{-0.25}$).

The high-precision TESS space photometry allows us to exploit not only one but two age estimators for open clusters, namely asteroseismology and gyrochronology. These two methods are particularly valuable independent age estimators because many of the young clusters and stellar associations discovered with \emph{Gaia} do not host evolved stars, which limits the capacity of isochrone fitting. All three age estimates for UBC\,1 are similar: the range of 150\,Myr to 300\,Myr is simultaneously compatible with isochrone fitting, g-mode asteroseismology and gyrochronology.

Despite the good agreement of the age-dating estimates, the overall systematic uncertainty remains large
due to the choice of different input physics for the isochrones and due to the limited yield of asteroseismology. For the asteroseismic age-dating, we could use only one measurement of $\Pi_0$ for a single
cluster member. It would be of great interest to probe whether all pulsators in the cluster have a compatible asteroseismic age and to achieve a proper age spread from ensemble asteroseismic modelling. Our work is a successful proof-of-concept study showing that even one pulsator with only a few identified g modes allows for a drastically improved age estimate compared to the use of its surface quantities $\log L/L_\sun$,  $T_{\rm eff}$, and $\log\,g$ measured by {\it Gaia}.

Currently, the gyrochronal age estimate is the most precise for UBC\,1 because it is the least model dependent. However, with more and more open clusters becoming available for g-mode asteroseismology thanks to TESS, we can start building a cluster modelling methodology that will eventually lead to an empirical age-ranked sequence based on $\Pi_0$ and $f_{\rm rot}$ estimates (and possibly other asteroseismic observables).
Indeed, with the increasing amount of space photometry from the ongoing TESS mission and the future PLATO \citep{Rauer2014} mission, we enter a golden era of g-mode open-cluster asteroseismology. An age-ranked sequence of (g-mode) asteroseismic observables would enable true calibrations of asteroseismic models to leverage their full potential.

\begin{acknowledgements}
    We are grateful to Aaron Dotter for the suggestion of using rotating isochrones and for further discussions. We thank the anonymous referee for the very useful review.
    The research leading to these results has received funding from the KU\,Leuven Research Council (grant C16/18/005: PARADISE).
CA also acknowledges financial support from the Research Foundation Flanders (FWO) under grant K802922N (Sabbatical leave).
    This research has made use of NASA's Astrophysics Data System Bibliographic Services and of the SIMBAD database and the VizieR catalogue access tool, operated at CDS, Strasbourg, France.
    This publication makes use of data products from the Wide-field Infrared Survey Explorer, which is a joint project of the University of California, Los Angeles, and the Jet Propulsion Laboratory/California Institute of Technology, and NEOWISE, which is a project of the Jet Propulsion Laboratory/California Institute of Technology. WISE and NEOWISE are funded by the National Aeronautics and Space Administration.
    This work has made use of data from the European Space Agency (ESA) mission \emph{Gaia} (\url{https://www.cosmos.esa.int/gaia}), processed by the \emph{Gaia} Data Processing and Analysis Consortium (DPAC, \url{https://www.cosmos.esa.int/web/gaia/dpac/consortium}). Funding for the DPAC has been provided by national institutions, in particular the institutions participating in the \emph{Gaia} Multilateral Agreement.
    This paper includes data collected by the TESS mission, which are publicly available from the Mikulski Archive for Space Telescopes (MAST).
    \newline
    \textbf{Software:}
    This research made use of \texttt{Astropy}, a community-developed core \texttt{Python} package for Astronomy \citep{2013A&A...558A..33A} and \texttt{Lightkurve}, a \texttt{Python} package for Kepler and TESS data analysis \citep{2018ascl.soft12013L}.
    This work made use of \texttt{Topcat} \citep{2005ASPC..347...29T}.
    This work made use of \texttt{ColorBrewer2} \url{http://www.ColorBrewer2.org}.
    This research made use of the following \texttt{Python} packages:
    \texttt{astroquery} \citep{2019AJ....157...98G};
    \texttt{corner} \citep{2016JOSS....1...24F};
    \texttt{IPython} \citep{ipython};
    \texttt{MatPlotLib} \citep{Hunter:2007};
    \texttt{NumPy} \citep{numpy};
    \texttt{Pandas} \citep{pandas};
    \texttt{SciPy} \citep{scipy};
    \texttt{seaborn} \citep{Waskom2021}

\end{acknowledgements}

\bibliographystyle{aa} 
\bibliography{UBC1.bib} 

\begin{appendix}
\section{ADQL query}
\label{app:adql}
For the membership analysis, we used the following ADQL query to download the \emph{Gaia}~DR3 data from ESA \emph{Gaia} archive at \url{https://gea.esac.esa.int/archive/}.

\begin{verbatim}
SELECT
g.source_id, g.ra, g.dec,
g.parallax, g.parallax_over_error,
g.distance_gspphot, g.pmra, g.pmdec,
g.radial_velocity, g.radial_velocity_error,
g.phot_g_mean_mag, g.bp_rp, g.g_rp,
g.phot_bp_mean_mag, g.phot_rp_mean_mag,
g.phot_g_mean_flux_over_error,
g.phot_bp_mean_flux_over_error,
g.phot_rp_mean_flux_over_error

FROM gaiadr3.gaia_source as g

WHERE

-- Use 15 deg radius around UBC 1
CONTAINS(
    POINT('ICRS', g.ra, g.dec),
    CIRCLE('ICRS', 288.0, 56.83, 15)
    )=1

-- Distance, proper motion, and magnitude limits
AND
    (g.parallax >= 2 AND
     g.parallax <= 5 AND
     g.pm<=30 AND
     g.phot_g_mean_mag <= 18)

\end{verbatim}

\section{Photometric binaries}
\label{app:binarity}
\begin{figure*}
    \includegraphics[width=\textwidth]{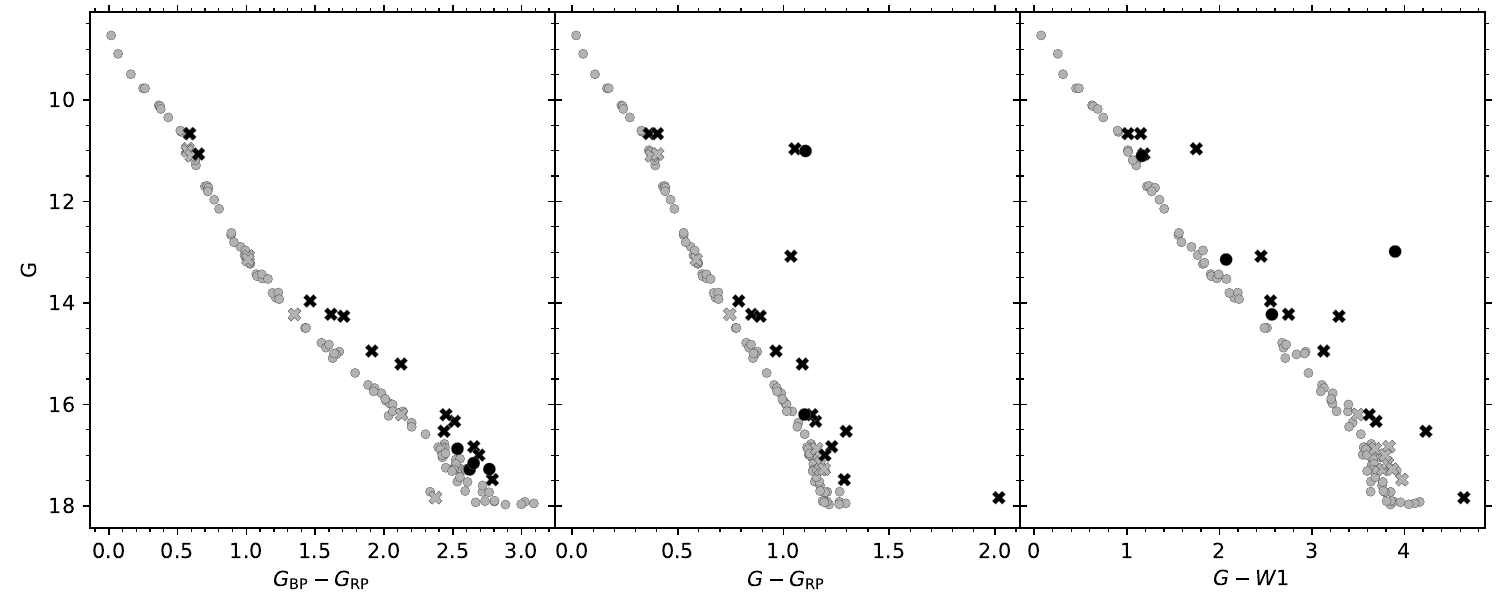}
    \caption{Colour-magnitude diagrams used to identify photometric binaries. In each panel, photometric binaries in the respective colour are highlighted. Crosses mark photometric binaries identified also in other colours, while highlighted circles are photometric binaries unique to this colour. Grey circles denote other members. For simplicity, we do not show binaries identified through other means.}
    \label{fig:phbinCMDs}
\end{figure*}

Our analysis of binarity includes photometric binaries that we identified through multi-colour photometry. The CMDs in Fig.~\ref{fig:phbinCMDs} show the cluster members in \emph{Gaia} $G$ against three different colours based on \emph{Gaia} and WISE observations. In each panel, we highlight the photometric binaries in this colour. We note that the photometry was not dereddened for this analysis because the reddening towards UBC\,1 is very small.

\section{Posterior distributions}
\begin{figure}
    \includegraphics[width=\columnwidth]{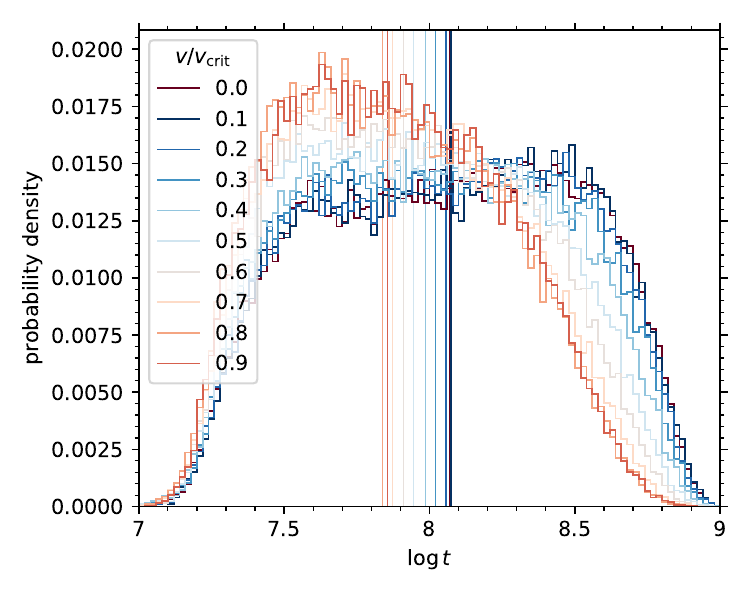}
    \caption{Posterior distribution of the open cluster age ($\log t$) for isochrones with different rotation rates. The dependence of the maximum likelihood age with the rotation rate is well visible. The vertical lines indicate the median values the distributions.}
    \label{fig:agedistrib}
\end{figure}

The posterior distributions of the isochrone fitting in Sect.~\ref{sec:isoage} have a strong trend in age with rotation rate. To illustrate this trend and to provide additional information to Fig.~\ref{fig:rotisocomp}, we show these distributions for all ten considered rotation rates in Fig.~\ref{fig:agedistrib}. The extinction is better constrained and depends only slightly on the rotation rate. Its posterior distribution is therefore not shown. We provide the numerical values for the median and maximum likelihood of both parameters in Table~\ref{tab:isopara}.

\begin{table}
    \caption{Isochrone parameters for each isochrone shown in Fig.~\ref{fig:rotisocomp}.}
    \label{tab:isopara}
    \begin{tabular}{rrrrr}
        \hline
        \hline
        $v/v_\mathrm{crit}$ & \multicolumn{2}{r}{median values} & \multicolumn{2}{r}{max. likelihood}  \\
         & $\log t$ & $A_V$ & $\log t$ & $A_V$\\
         &&(mag)&&(mag)\\
        \hline
        0.0 & 8.09 & 0.09 & 8.37 & 0.07\\
        0.1 & 8.09 & 0.09 & 8.45 & 0.09\\
        0.2 & 8.07 & 0.09 & 8.23 & 0.06\\
        0.3 & 8.03 & 0.09 & 8.25 & 0.03\\
        0.4 & 8.01 & 0.08 & 8.19 & 0.05\\
        0.5 & 7.97 & 0.08 & 8.07 & 0.04\\
        0.6 & 7.91 & 0.07 & 7.54 & 0.04\\
        0.7 & 7.87 & 0.06 & 7.58 & 0.02\\
        0.8 & 7.85 & 0.06 & 7.70 & 0.01\\
        0.9 & 7.87 & 0.05 & 7.58 & 0.00\\
        \hline
    \end{tabular}
\end{table}

\section{Power spectra of cluster members}
\label{app:othervariables}

In this section, we display the frequency spectra of variable intermediate mass stars, that do not show a clear pulsation characteristic but are still variable at $f>1$\,d$^{-1}$. Their characteristics are briefly discussed in the main text. We omit the frequency spectra for UBC1-68 and UBC1-97 as they are featureless for $f>1$\,d$^{-1}$. For the same reason, we only show the frequency range $0\le f\le10$\,d$^{-1}$ in the figures.

\begin{figure*}
    \includegraphics[width=0.49\textwidth]{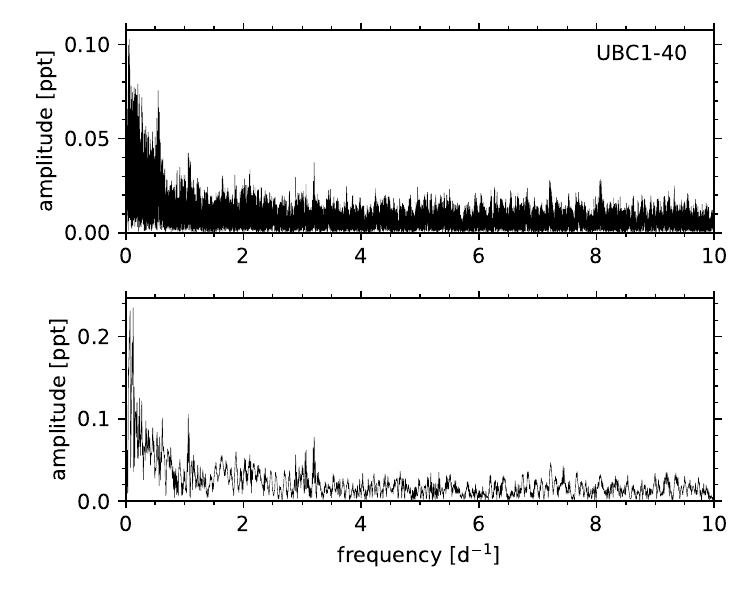}\quad
    \includegraphics[width=0.49\textwidth]{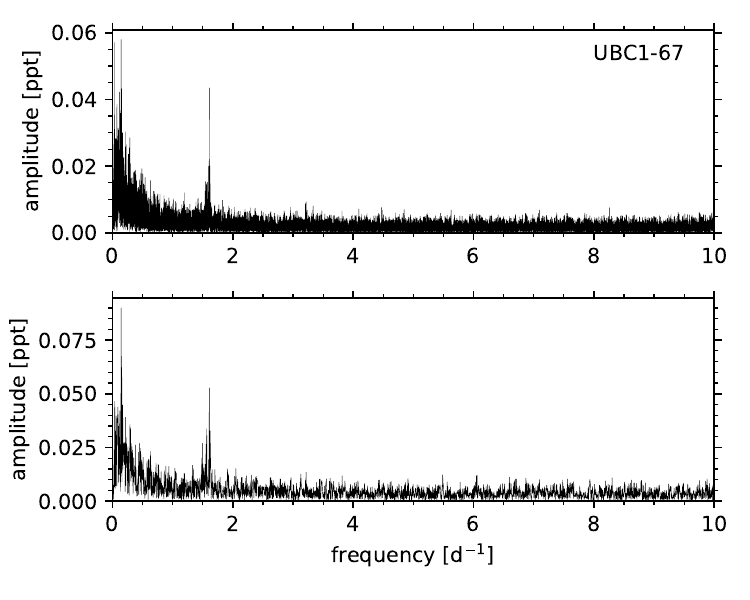}\\
    \includegraphics[width=0.49\textwidth]{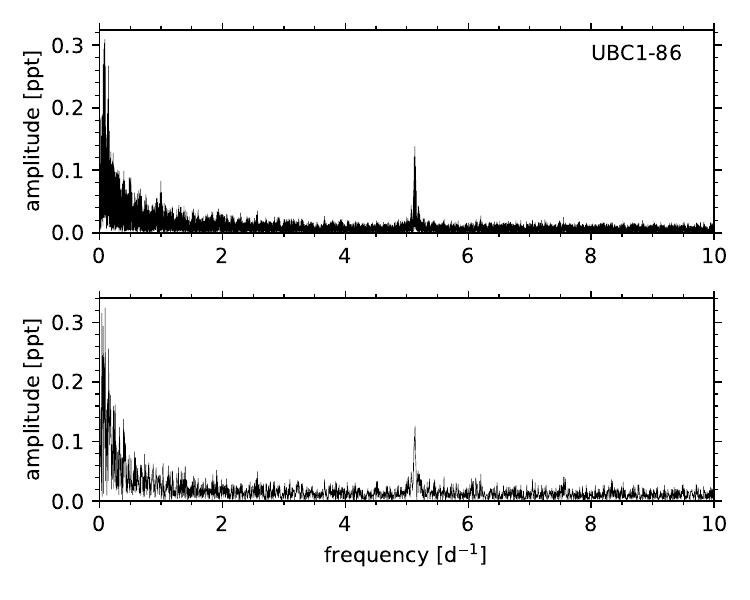}\quad
    \includegraphics[width=0.49\textwidth]{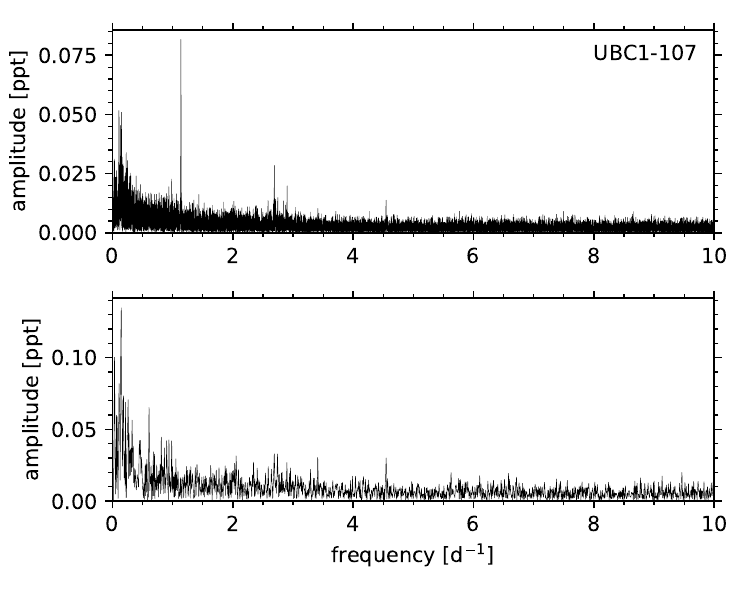}
    \caption{Frequency spectra for variable intermediate mass UBC\,1 members. The observations include 18 sectors for UBC1-40 (\emph{top left}), 21 sectors for UBC1-67 (\emph{top right}), 21 sectors for UBC1-86 (\emph{bottom left}), and 27 sectors for UBC1-107 (\emph{bottom right}). For each star the top panel shows the periodogram based on the full TESS data set, while the lower panel is based only on the 200\,s data from Sectors $56-60$.}
    \label{fig:pg40}
\end{figure*}

\section{Light curves of cool star rotators}
\label{app:LCs}
\begin{figure*}
    \includegraphics[width=\textwidth]{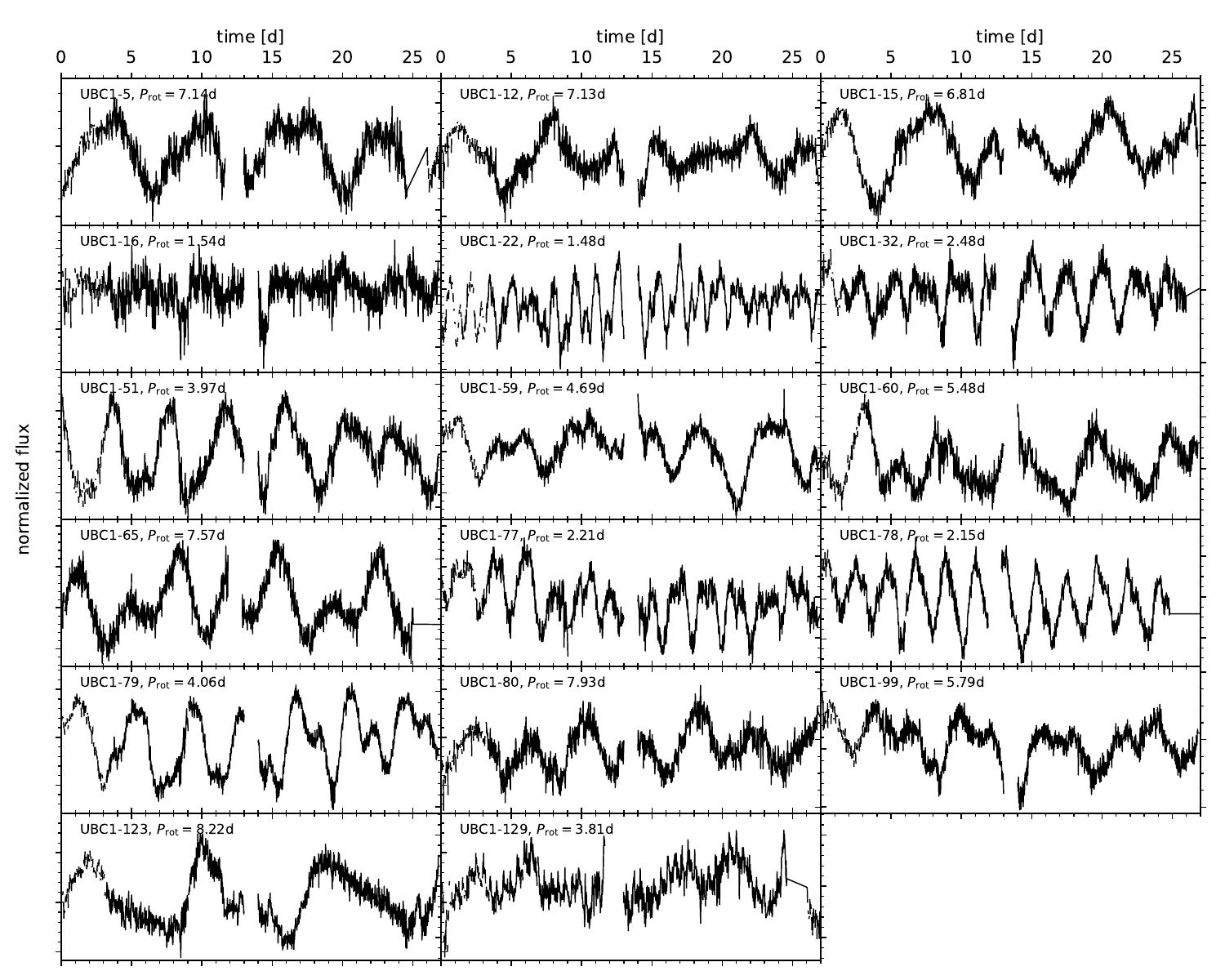}
    \caption{Light curves for the 17 identified cool star rotators in UBC\,1. Most stars show very clear rotational modulation. For each star, we show only the first available TESS sector to enable a visual estimate of the rotation period.}
    \label{fig:rotLC}
\end{figure*}

In Fig.~\ref{fig:rotLC}, we show light curves of the 17 identified cool star rotators. For each star, we display only the first available TESS Sector and indicate the rotation period at the top. We note that these rotation periods are derived from the full light curves.

\end{appendix}

\end{document}